\documentclass[a4paper, 11pt]{article}	
\usepackage{url}
\usepackage{comment}
\usepackage{acronym}
\usepackage{natbib}
\usepackage{graphicx}
\usepackage{eurosym}
\begin{document}

\acrodef{TSP}{Travelling Salesman Problem}
\acrodef{MTSP}{Multiple Travelling Salesmen Problem}
\acrodef{CO}{Centralised Organisation}
\acrodef{DO}{Decentralised Organisation}
\acrodef{CA}{Central Authority}
\acrodef{MILP}{Mixed-Integer Linear Programming}
\acrodef{RFP}{Request For Proposals}
\newcommand{\TODO}[1]{{\bf #1}}

\title{
Cost of selfishness in the allocation of cities in the\\
Multiple Travelling Salesmen Problem}
\author{Thierry Moyaux$^1$ \& Eric Marcon$^2$\\
\small 1. Univ Lyon, INSA-Lyon, DISP, EA 4570, F-69621, France.\\
\small Corresponding author.\\
\small 2. Univ Lyon, INSA-Lyon, UJM-Saint Etienne, DISP, EA 4570, F-69621, France.}
\maketitle

\abstract
The decision to centralise or decentralise human organisations requires quantified evidence but little is available in the literature. We provide such data in a variant of the \acf{MTSP} in which we study how the allocation sub-problem may be decentralised among selfish selfmen.
Our contributions are
(i) this modification of the \ac{MTSP} in order to include selfishness,
(ii) the proposition of organisations to solve this modified \ac{MTSP}, and
(iii) the comparison of these organisations.
Our 5 organisations may be summarised as follows:
(i) \sf{OptDecentr} is a pure \ac{CO} in which a \ac{CA} finds the best solution which could be found by a \ac{DO},
(ii) \sf{Cluster} and (iii) \sf{Auction} are \ac{CO}/\ac{DO} hybrids, and
(iv) \sf{P2P} and (v) \sf{CNP} are pure \ac{DO}.
Sixth and seventh organisations are used as benchmarks:
(vi) \sf{NoRealloc} is a pure \ac{DO} which ignores the allocation problem, and
(vii) \sf{FullCentr} is a pure \ac{CO} which solves a different problem, $viz.,$ the traditional \ac{MTSP}.
Comparing the efficiency of pairs of these mechanisms quantify the price of decentralising an organisation.
In particular, our model of selfishness in \sf{OptDecentr} makes the total route length 30\% (respectively, 60\%) longer with 5 (respectively, 9) salesmen than the traditional \ac{MTSP} in \sf{FullCentr} when the computation time is limited to 30 minutes.
With this time limit, our results also seem to indicate that the level of coercion of the \ac{CA} impacts more the total route length than the level of centralisation.

\smallskip
\noindent \textbf{Keywords.} \acf{CO}; \acf{DO}; Selfishness; \acf{MTSP}.


%
%
%
%
\section{Introduction}

Everybody has an opinion about the centralisation of decision. We believe that this opinion depends too much on intuition rather than quantified evidence. In addition, several criteria exist to compare organisations. For instance, \acf{CO} is often said to find the optimum solution (or, at least, the best possible solution), while \acf{DO} is more reactive. Two incomparable metrics are opposed here, namely, quality of the solution found and computation speed.
Like comparisons of capitalism and socialism \citep{ellman89}, numerical comparisons of \ac{CO} and \ac{DO} rely on comparisons of specific organisations instantiating these levels of centralisation. In other words, it is difficult to provide numerical evidence of the efficiency of \ac{CO} and \ac{DO} in general, hence particular mechanisms of these approaches are compared.
In this article, we call ``mechanism'' an instance of an ``organisation'', and an ``organisation'' is an instance of an ``approach''.
The possible ``approaches'' are pure \ac{CO} and \ac{DO}, and \ac{CO}/\ac{DO} hybrids.
Let us illustrate these three terms:
The \ac{CO}/\ac{DO} hybrid approach can be instantiated as an ``organisation'' with one or several (coercive) dispatchers, and/or (non-coercive) mediators such as an auctioneer, or peer-to-peer negotiation, etc.
Next, the ``organisation'' with an auctioneer can be instantiated as a ``mechanism'' as either an English (multiple-shot first-price) or Vickrey (single-shot second-price) auction.

We think that the choice of a more or less centralised organisation is a question too often addressed by political discussions as a consequence of the lack of numerical evidence.
This research question is important because the efficiency of organisations may be greatly improved by selecting the appropriate decision organisation. In fact, defenders of \ac{CO} (respectively, \ac{DO}) spends much effort in developing their mechanism, while they may obtain much better performances by introducing some features of \ac{DO} (respectively, \ac{CO}) --- ``If your only tool is a hammer then every problem looks like a nail'', as the saying goes. 
Such a choice between various organisations requires numerical data about these organisations, such as the aforementioned quality of the solution and computation time.
With such data, an organisation could be designed depending on its constraints; for instance, choosing the number of hierarchical levels and the appropriate organisation on each level depending on the time available to make a decision.
In particular, we believe that our work will shed light on the organisation of the Physical Internet \citep{sarraj14} since this project aims at connecting \ac{CO} logistics networks in a \ac{DO} way.

The literature provides little quantified comparisons of \ac{CO} and \ac{DO}, as noted in a review co-written by one of us~\citep{moyaux12}. In addition to the papers referenced in this review, we now outline others which compare \ac{CO} and \ac{DO} in applications to logistics.
In this area, the scarcity of quantified comparisons has also been noted \cite[p.~60]{mes07}.
Similarly, \cite{davidsson05-bis} regret the few quantitative comparisons of \ac{DO} approaches with \ac{CO}.
One of the earliest such comparisons in logistics is a work by \cite{fisher95, fisher96} who propose an extension of \cite{smith80}'s Contract Net Protocol (CNP) in which task decomposition is decentralised in order to solve a static Vehicle Routing Problem with Time Windows. Their experiment shows that their protocol finds a total route length between 3\% and 74\% worse than the optimal and is thus comparable to heuristics from Operations Research. Then, they propose an improvement which reduces these figures by about~12\%.
Since routing problems are NP-hard, the optimal route is often unknown and some studies compare the results of \ac{DO} with those of approximate \ac{CO} heuristics. In this context, \cite{mes07} report that their Vickrey auction (\ac{DO}) performs as good as or even better than centralised heuristics to solve a dynamic Pickup and Delivery Problem with Time Windows (PDPTW).
For the same problem, \cite{mahr10} also compare a Vickrey auction (\ac{DO}) to the approximate solution found by CPLEX in a few 30-second intervals (\ac{CO}). Their experimentation shows that the auction outperforms CPLEX when service time duration is highly uncertain.
Next, \cite{vonLon15} compare a cheapest insertion heuristic (\ac{CO}) with an implementation of \cite{fisher95}'s Contract Net Protocol (\ac{DO}). Their results seem to indicate that \ac{DO} outperforms \ac{CO}.
\cite{glaschenko09} have implemented a real-time multiagent scheduler for a taxi company in London, UK. The benefit of their tool is distributed between the drivers and the company, resulting in an increase of driver wages by 9\%.
\cite{karmani07} use a market-based approach to solve a capacitated version with multiple depot of the \ac{MTSP}. Their experiment shows that their approach scales to thousands of cities and hundreds of salesmen with a total route length quite close to the optimum.
Later, \cite{kivelevitch13} were inspired by this approach to propose a distributed meta-heuristic. 
The total length of the routes are within 1\% of the optimal in 90\% of the tested cases. More precisely, the median total length of the route is 1.7\% higher than the optimum and 4.8\% higher in the worst of the tested cases, but the median run time is 8 times faster than CPLEX.
Nevertheless, this approach uses agents with no autonomy at all ($e.g.,$ an agent can take one or all cities from another agent without permission), hence we think it is more related to parallel algorithms \citep{talbi06} than to the making of decisions in human organisations with selfish agents as we investigate.
Quite similarly, others solve logistic problems ($e.g.,$ \ac{TSP} by \cite{xie09} and \cite{hanna09}) with agents who, again, lack the human characteristics of selfishness and cannot hence be used to design organisations as well.


The goal of this article is to compare \ac{CO} with \ac{DO} by measuring the efficiency of several mechanisms on various levels of centralisation to solve a modified \acf{MTSP} (\ac{MTSP} is the Vehicle Routing Problem without capacities) such that the selfishness of the salesmen is taken into account.
It is important to notice that our salesmen are selfishness because we want to investigate human organisations.
This article has the following contributions:
\begin{enumerate}
	\item \emph{Conceptual contributions}: \ac{CO} and \ac{DO} are different by nature because, for example, the social welfare ($i.e.,$ utility function of the group) in \ac{CO} may have no trivial relationship with the individual utility functions in \ac{DO}. Game Theory proposes models and tools to study \ac{CO} and \ac{DO}, $e.g.,$ the Prisoner's Dilemma shows that \ac{DO} may find a solution ($i.e.,$ Nash equilibrium) which is Pareto dominated by the solution found by \ac{CO}. Taking this game-theoretic background into account, we modify the \ac{MTSP} by adding features representing the selfishness of the salesmen such that our modified \ac{MTSP} can be solved in more or less centralised organisations. Selfishness means that a salesman maximises his\footnote{We always use ``him'' for a salesman and ``her'' for the \acf{CA}.} own utility in a selfish manner without paying attention to the social welfare of the community of salesmen.

Subsection~\ref{modifiedMTSP} introduces our \ac{MTSP} constrained by 1-1 exchanges. More precisely, the salesmen ``own'' an initial endowment of cities, and they use a mechanism to modify this initial allocation of cities by exchanging one city against one city exclusively, which we refer to as ``1-1 exchanges''. This corresponds to our modified \ac{MTSP}. Section \ref{section-discussion} discusses other possibilities to obtain a pair of \ac{CO} and \ac{DO} versions of a problem.

	\item \emph{Technical contributions}: As said in the above literature review, no previous work compares more than two kinds of organisations to solve a same location and routing problem. Our organisations for the exchange of cities are inspired by the classes of coordination proposed in \cite[p.~131]{frayret02} as an extension to \cite{mintzberg78}.
To be precise, this paper studies the following organisations which have various numbers of hierarchical levels, coercion levels and number of rounds of interactions, as summarised in Figure~\ref{organisations.eps}:
\begin{enumerate}
	\item[1.] {\sf OptDecentr} makes the \acf{CA} find both (i) the best allocation of cities to salesmen with the constraint of 1-1 exchanges and (ii) for each salesman, the shortest path visiting all the cities allocated to this salesman -- this is a \ac{CO} mechanism which looks for the best solution which may be found by the following five \ac{DO} mechanisms;
	\item[2.] \sf{Cluster} makes the \ac{CA} allocate groups of neighbouring cities to the salesmen with the constraint of 1-1 exchanges, then every salesman locally solves a \ac{TSP} with his allocated cities;
	\item[3.] \sf{Auction} is similar to \sf{Cluster} except that the \ac{CA} is less coercive since she plays the role of an auctioneer;
	\item[4.] \sf{CNP} (Contract Net Protocol) has no \ac{CA} and every salesman plays the role of an auctioneer;
	\item[5.] \sf{P2P} also has no \ac{CA} and relies on bilateral negotiations.
\end{enumerate}

Finally, two additional organisations are used as benchmarks:
\begin{enumerate}
	\item[6.] \sf{NoRealloc} assumes no reallocations of the cities among the salesmen who only solve a \ac{TSP} on their initial endowment;
	\item[7.] \sf{FullCentr} is the same as \sf{OptDecentr} without the constraint of 1-1 exchanges, that is \sf{FullCentr} is the only mechanism that solves the traditional \ac{MTSP}.
\end{enumerate}
\begin{figure}[t]
    \begin{center}
    \centering
    \includegraphics[width=14.5cm]{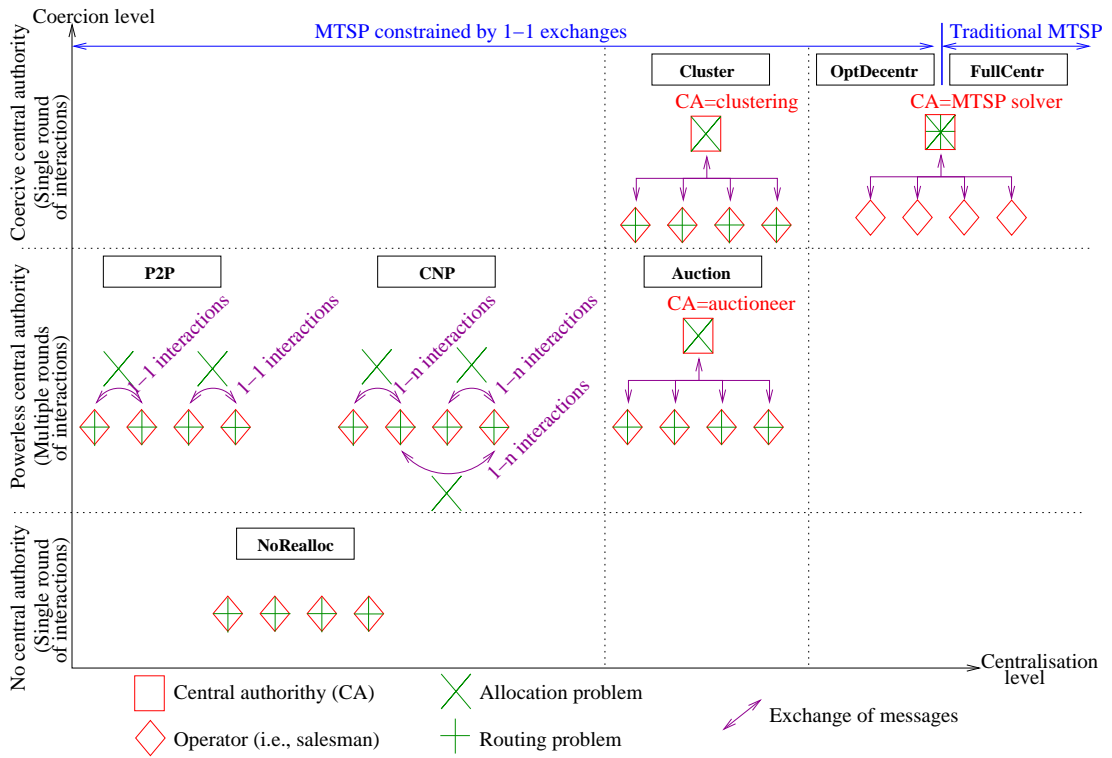}
    \caption{Overview of the seven organisations compared.}
    \label{organisations.eps}
    \end{center}
\end{figure}

Since our experimentation simulates more agents than the number of cores of the CPU in our computers, we perform only sequential simulations, then we infer what would have happened in real life with computations carried out in parallel.

As said above, we call ``mechanism'' an instance of an ``organisation''. However, this article uses both terms interchangeably because it deals with one mechanism per organisation only. In fact, each of the above organisations may be instantiated in mechanisms different from our descriptions in Section~\ref{section-organisations}, as shall be discussed in Section~\ref{section-discussion}.

Finally, our mechanisms instantiating the above first five organisations are new, since they solve the \ac{MTSP} constrained by 1-1 exchanges which we have never seen in the literature. On the contrary, benchmarks \sf{NoRealloc} and \sf{FullCentr} are not new as they solve a classic \ac{MILP} formulation of \ac{TSP} and \ac{MTSP}.

	\item \emph{Numerical contributions}: \cite{davidsson07} compare four mechanisms on a same problem and we do not know of any work with more mechanisms. In comparison with this work, our work both consider more mechanisms and present many more experimental results. In fact, our results are robust since we compare the efficiency of the mechanisms when they solve the same instance among 130 for given numbers of salesmen and clients, and we present the fifth and ninth deciles obtained by 130 instances.
\end{enumerate}

The outline of this paper is as follows.
Section \ref{section-framework} presents our framework, $i.e.,$ our \ac{MTSP} constrained by 1-1 exchanges, next hypotheses about how to carry out a fair comparisons of the mechanisms.
Section \ref{section-organisations} details the seven mechanisms.
Section \ref{duration} shows how the duration of the parallel computations of up to 10 agents (9 salesmen + 1 \acf{CA}) is inferred from the sequential computations in our experimentation on one of our computers with a 4-core CPU.
Section \ref{experimentation} presents the numerical results of this experimentation.
Section \ref{section-discussion} discusses these results, next how other variants of \ac{MTSP} may be introduced taking the selfishness of the salesmen into account.
Section \ref{Conclusion} concludes.

\section{Framework}
\label{section-framework}

This section first recalls the formulation of the traditional \ac{MTSP} solved by Mechanism \sf{FullCentr}, next our \ac{MTSP} constrained by 1-1 exchanges of cities which is solved by the other six mechanisms.
We assume that both problems have $n$ cities to be visited by one of $m$ salesmen, and $d_{ij}$ is the Euclidean distance between Cities $i$ and~$j$. City $i=0$ is the depot shared by all salesmen.

\subsection{Traditional \ac{MTSP}: \sf{FullCentr}}
\label{traditionalMTSP}

The traditional \ac{MTSP} in Equations \ref{eq-ECS-obj}-\ref{eq-ECS-var} uses the 2-index decision variable $x_{ij}$ such that $x_{ij}=1$ only if one of the $m$ salesmen goes from City $i$ to City~$j$.
The objective function in Equation \ref{eq-ECS-obj} minimises the total distance $U^\textrm{trad}$ travelled by all the salesmen.
Equation \ref{eq-ECS-flowOutDepot} (respectively, \ref{eq-ECS-flowInDepot}) ensures that $m$ salesmen leave (respectively, enter) the depot.
Similarly, Equation \ref{eq-ECS-flowOut} (respectively, \ref{eq-ECS-flowIn}) ensures that a single salesman leaves (respectively, enters) every city.
Equation~\ref{eq-ECS-subroute} is a constraint eliminating sub-routes by the method of node potentials proposed by \cite{miller60} in which decision variable $p_i$ counts the number of cities visited by a salesman before he visits City $i$.
\begin{eqnarray}
\min U^\textrm{trad}=	&\sum_{i=0}^{n-1} \sum_{j=0, j\neq i}^{n-1} d_{ij} x_{ij}		\label{eq-ECS-obj}\\
	&\sum_{j=1}^{n-1} x_{0j}=m			&					\label{eq-ECS-flowOutDepot}\\
	&\sum_{j=0, j\neq i}^{n-1} x_{ij}=1		&1\leq i < n				\label{eq-ECS-flowOut}\\
	&\sum_{i=0, i\neq j}^{n-1} x_{ij}=1		&1\leq j < n				\label{eq-ECS-flowIn}\\
	&\sum_{i=1}^{n-1} x_{i0}=m			&					\label{eq-ECS-flowInDepot}\\
	&p_i - p_j + (n-1).x_{ij} \leq n-2		&1\leq i \neq j < n			\label{eq-ECS-subroute}\\
	&p_i\in \Re^+					&1\leq i < n				\label{eq-ECS-varU}\\
	&x_{ij}\in \{0,1\}				&0\leq i,j < n				\label{eq-ECS-var}
\end{eqnarray}

\subsection{Modified \ac{MTSP} (constrained by 1-1 exchanges)}
\label{modifiedMTSP}

We now explain how we modify this traditional \ac{MTSP} in order to include the individual selfishness of the salesmen such that a \ac {DO} can also solve our modified problem. The following explanation calls $U^\textrm{trad}$ the social welfare in the traditional \ac{MTSP} ($i.e.,$ the total route length in Eq.~\ref{eq-ECS-obj}), and $U^\textrm{mod}$ the social welfare and $u^\textrm{mod}_k$ Salesman $k$'s utility in the modified \ac{MTSP}.
We make the following assumptions:
\begin{enumerate}
	\item \sf{Hyp. 1} \emph{-- Individual utilities measure the travelled distance}: We define $u^\textrm{mod}_k$ as the distance travelled by Salesman $k$ and $U^\textrm{mod}=\sum_k u^\textrm{mod}_k$ as the distance travelled by all. Of course, the advantage of such a hypothesis is that $U^\textrm{mod}=U^\textrm{trad}$.

	The drawback is that salesmen all want to get rid of their cities, but no one accepts cities from other salesmen. Consequently, such choices of $u^\textrm{mod}_k$ and $U^\textrm{mod}$ require the following two assumptions which operate together.
	\item \sf{Hyp. 2} \emph{-- Salesmen ``own'' an initial endowment of cities}: Since \sf{Hyp. 1} makes our distance-minimiser salesmen all want to reduce the number of cities allocated to them, we cannot assume that they will ``fight'' in order to obtain cities from a shared pool. Instead, we assume that every Salesman $k$ initially ``owns'' some cities, and he tries to change this initial allocation by exchanging cities whenever such an exchange reduces his individual distance~$u^\textrm{mod}_k$.
	\item \sf{Hyp. 3} \emph{-- Only 1-1 exchanges of cities are possible}: Let us study what happens when a Salesman $k$ gives $n$ cities to Salesman $k'$ and receives $n'$ cities from $k'$ in a round of interaction. All exchanges with $n=n'$ are rational for both salesmen when they reduce both $u^\textrm{mod}_k$ and $u^\textrm{mod}_{k'}$, as shown in the next paragraph. This constraint of $n-n$ exchanges solves the above problem caused by selfishness which makes salesmen want to give but not receive cities. In this article, we only consider the case $n=n'=1$; Cases $n=n'>1$ are discussed in Section~\ref{section-discussion}.

	Let us detail why $n<n'$ cannot be accepted by $k'$ (the case $n>n'$ is similar). Without knowing the cities allocated to Salesman $k'$ (such a list of clients is private in \ac{DO}), Salesman $k$ will have trouble convincing $k'$ to accept more than $n'=n$ city in an interaction round. In fact, a consequence of the triangular inequality is that accepting $n$ (respectively, $n+1$, $n+2$, etc.) cities while giving $n-1$ (respectively, $n$, $n+1$, etc.) increases more $u^\textrm{mod}_{k'}$ than accepting the same number of cities as the number given (except when a city exactly lies on the optimal route -- which is known by $k'$ but ignored by~$k$). As a result, even when $k$ proposes to give $n$ close cities in exchange for $n'$ cities, $k'$ would not be rational to accept $n'<n$.
\end{enumerate}

The traditional \ac{MTSP} is the problem faced, for example, by a home health care service in which all nurses are employees and have hence no individual utility function to optimise.
In contrast, our modified \ac{MTSP} corresponds to the problem faced by a private nurse association in which the nurses have an initial endowment of patients (the patients either make an appointment to their nurse or are provided by a physician) and the association is a place in which the nurses may exchange patients when this is mutually beneficial. In this modified problem, every private nurse has an individual utility function which they optimise.

\subsection{Hypothesis \sf{Hyp.~4} about the computation time}

In this article, we also make Assumption \sf{Hyp.~4} which states that we only take account of the computation time of the \ac{MILP} solver, namely CPLEX. We choose to make this assumption as a solution to the problem of comparing various organisations implemented in various environments and languages. In fact, \sf{Hyp.~4} forbids the comparison of the duration of the operation of a mechanism which mostly uses CPLEX with the duration of another which simulates its interactions in AnyLogic\footnote{The AnyLogic model and the outcomes of the experiments will be published on \url{www.github.com} after acceptation of this article for publication.} which runs in Java.
Because of this hypothesis, we ignore the duration of anything not computed by CPLEX, such as:
\begin{itemize}
	\item \emph{Messages travel instantaneously}: \ac{DO} implies interactions which are simulated by AnyLogic, not CPLEX. Hence, we ignore the travelling time of the messages exchanged.
	\item \emph{Some techniques are not possible}: We cannot use multiagent techniques, such as BDI (Belief-Desire-Intention) architecture \citep{caballero11} or reinforcement learning \citep{gunady14}, which may decrease the performance of our \ac{DO} organisations. Similarly, Organisation \sf{Cluster} cannot use the k-means algorithm, but a \ac{MILP} formulation usable by CPLEX. Likewise, the salesmen locally solve a \ac{TSP} in several of our organisations with CPLEX, but they cannot use Concorde\footnote{\url{http://www.math.uwaterloo.ca/tsp/concorde/index.html}} while it is often seen as the most efficient.
\end{itemize}


\section{Allocation mechanisms}
\label{section-organisations}

We now describe our six mechanisms/organisations to solve the \ac{MTSP} constrained by 1-1 exchanges.
Figure~\ref{organisations.eps} shows an overview in which we can see, for example, that \sf{OptDecentr} is more centralised than \sf{Cluster} and \sf{Auction}, and \sf{Cluster} has a more coercive \ac{CA} than \sf{Auction}.
This figure also points out that \sf{OptDecentr} is the most centralised in the sense that the \ac{CA} solves both the allocation and all routing problems while the other organisations let the salesmen locally solve a \ac{TSP} on their allocated cities. Some organisations operate in a single round while others need more interactions. Finally, this figure recalls that {\sf FullCentr} solves a problem different than the other mechanisms.
Each subsequent subsection details a mechanism.

\subsection{Pure \ac{DO}: \sf{NoRealloc}, \sf{P2P} and \sf{CNP}}
We first present \sf{NoRealloc} since its \ac{MILP} formulation of \ac{TSP} is both used in \sf{P2P}, \sf{Cluster} and \sf{Auction}, and the base of the \ac{MILP} formulation of the \ac{MTSP} constrained by 1-1 exchanges in \sf{OptDecentr}.

%
%
\subsubsection{\sf{NoRealloc}}
As said above, \sf{NoRealloc} ignores the allocation problem and lets every salesman find the shortest route leaving the depot, visiting the $N$ cities\footnote{\label{fnote} $N$ may be different between salesmen in the two variants of \ac{TSP} shown in this article. We do not use $N_k$ because $N$ is a local variable for every salesman-agent~$k$.} allocated to him in the initial endowment and returning to the depot. There is a single round in which each of the $m$ salesmen solves the \ac{TSP} in Equations~\ref{eq-TSP-obj}-\ref{eq-TSP-var}. This formulation uses the 2-index decision variable $x_{ij}$ which equals one only if the considered salesman goes from City $i$ to City~$j$.
\begin{eqnarray}
\min	&\sum_{i=0}^{N-1} \sum_{j=0, j\neq i}^{N-1} d_{ij} x_{ij}				\label{eq-TSP-obj}\\
s.t.	&\sum_{j=0, j\neq i}^{N-1} x_{ij}=1			&0\leq i < N			\label{eq-TSP-flowOut}\\
	&\sum_{i=0, i\neq j}^{N-1} x_{ij}=1			&0\leq j < N			\label{eq-TSP-flowIn}\\
	&p_i - p_j + N.x_{ij} \leq N-1				&1\leq i \neq j < N		\label{eq-TSP-subroute}\\
	&p_i\in \Re^+						&1\leq i < N			\label{eq-TSP-varU}\\
	&x_{ij}\in \{0,1\}					&(i,j)\in\{0,1, \ldots N-1\}^2	\label{eq-TSP-var}
\end{eqnarray}
With this notation, Equation \ref{eq-TSP-obj} is the same as Equation \ref{eq-ECS-obj} except that it minimises the route length of a single salesman and $n$ is thus replaced by $N$,
Equation \ref{eq-TSP-flowOut} (respectively, \ref{eq-TSP-flowIn}) is similar to Equations \ref{eq-ECS-flowOutDepot} and \ref{eq-ECS-flowOut} (respectively, \ref{eq-ECS-flowIn} and \ref{eq-ECS-flowInDepot}) except that it does not need to distinguish the case of the depot,
and Equation \ref{eq-TSP-subroute} is the same constraint of sub-route elimination as Equation~\ref{eq-ECS-subroute}.

%
%
\subsubsection{\sf{P2P}}
\begin{figure}
    \begin{center}
    \centering
    \includegraphics[width=13.5cm]{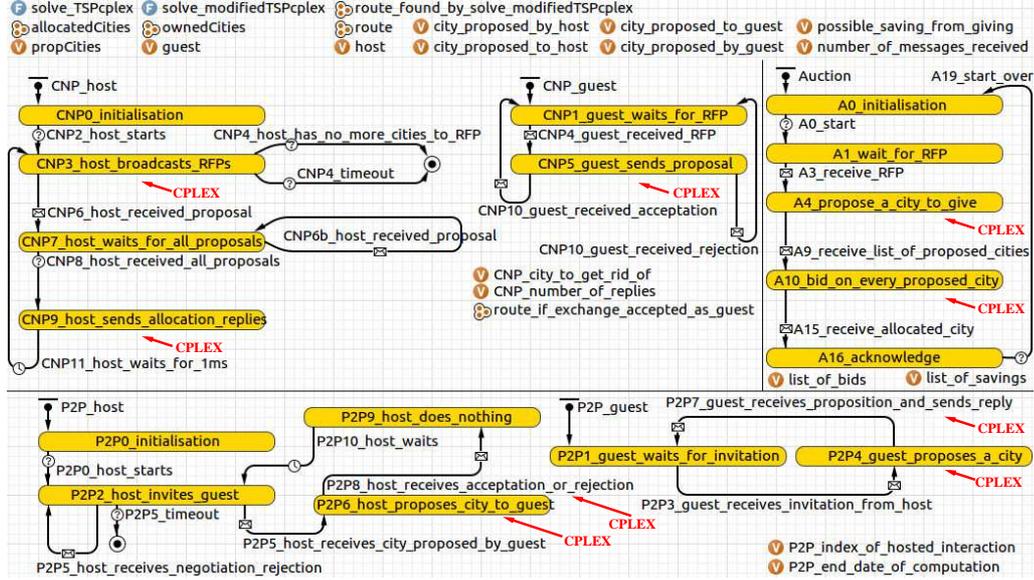}
    \caption{Implementation of the mechanisms in a salesman. (``CPLEX'' arrows point to transitions and states whose duration are taken into account.)}
    \label{xfig/salesman.eps}
    \end{center}
\end{figure}

Mechanism \sf{P2P} has several interaction rounds in which instances of \ac{TSP} or a derivative of \ac{TSP} are solved. The bottom of Figure~\ref{xfig/salesman.eps} shows that \sf{P2P} consists of two state charts which may run concurrently, namely, {\tt P2P\_host} and {\tt P2P\_guest} which replies to the former.
Therefore, every salesman may take part in two interactions simultaneously, the first as guest and the second as host. Please remember that every salesman operates his own copy of the state charts in Figure~\ref{xfig/salesman.eps}. The names of the states and transitions in both state charts all start with {\tt P2Pn\_action} where {\tt n} indicates their order of activation in a round and {\tt action} summarises the action performed.
\sf{P2P} uses the variables at the top of Figure~\ref{xfig/salesman.eps}. Each of these variables is a pointer to a city or salesman, except $propCities[k][i]$ that records previous interactions as a matrix of Booleans which are true only when the considered salesman has already proposed City $i$ to another Salesman~$k$ in order to prevent infinite loops. $allocatedCities$ (list of cities currently allocated to the considered salesman), $ownedCities$ (his initial endowment) and $route$ (similar to $allocatedCities$ but with the cities ordered according to the shortest route found by the mechanism in use, thus \sf{P2P} in this subsection) also are variables but AnyLogic shows them with a different icon because they are collections of objects.

We now detail the operation of \sf{P2P}. State {\tt P2P0\_initialisation} mainly sets all entries in $propCities[k][i]$ to $false$. AnyLogic always executes the first state in all state charts, and transition {\tt P2P0\_host\_starts} only fires when mechanism \sf{P2P} is selected. The first salesman whose {\tt P2P0\_host\_starts} fires first becomes the host in this round. (This state chart could generate more than one host at the same time, but we prevent this by setting AnyLogic to use only one thread, as detailed in Section~\ref{duration}.) Next, {\tt P2P2\_host\_invites\_guest} makes this host select a guest ($i.e.,$ the salesman with the lowest number of $false$ in $propCities[k][i]$) and send him an invitation.
The guest was waiting in state {\tt P2P1\_guest\_waits\_for\_invitation} and this message fires his transition {\tt P2P3\_guest\_receives\_invitation\_from\_host}.

Then, {\tt P2P4\_guest\_proposes\_a\_city} is the first call to CPLEX in this round of interaction. The guest solves the modified \ac{TSP} in Equations \ref{eq-p2p-guest-obj}-\ref{eq-p2p-guest-var} in order to propose a city to the host. The goal of this model is to find the city which should be removed from $allocatedCities$ such that the reduction of route length is maximised.\footnote{The problem in Equations \ref{eq-p2p-guest-obj}-\ref{eq-p2p-guest-var} finds the city which causes the largest increase in the route length. Instead of this single round, we may solve this problem in $N-1$ rounds in which the \ac{TSP} in Equations \ref{eq-TSP-obj}-\ref{eq-TSP-var} is solved with $N-1$ cities (the $i^\textrm{th}$ round without the $i^\textrm{th}$ city), then the shortest of the $N-1$ obtained route lengths indicates the city to be proposed. We have tested this method and seen that it takes more time than solving Equations \ref{eq-p2p-guest-obj}-\ref{eq-p2p-guest-var}.}
Equations \ref{eq-p2p-guest-obj}-\ref{eq-p2p-guest-var} modify the formulation of \ac{TSP} in Equations \ref{eq-TSP-obj}-\ref{eq-TSP-var} by adding binary decision variable $kept$ such that $kept_i=1$ for the $N-1$ cities to be kept and $kept_i=0$ for the city to be proposed to the host.\footnote{When this problem is solved by a salesman in Mechanism \sf{Auction}, the end of this sentence reads: ``$\ldots kept_i=0$ for the city to be proposed to the auctioneer''.}
\begin{eqnarray}
\min	&\sum_{i=0}^{N-1} \sum_{j=0, j\neq i}^{N-1} c_{ij} x_{ij}				\label{eq-p2p-guest-obj}\\
s.t.	&\sum_{j=0, j\neq i}^{N-1} x_{ij}=kept_i	&0\leq i < N				\label{eq-p2p-guest-flowOut}\\
	&\sum_{i=0, i\neq j}^{N-1} x_{ij}=kept_j	&0\leq j < N				\label{eq-p2p-guest-flowIn}\\
	&\sum_{i=1}^{N-1} kept_i = N-2			&0\leq i < N				\label{eq-p2p-guest-nb-of-cities-kept}\\
	&p_i - p_j + N.x_{ij} \leq N-1			&1\leq i \neq j < N			\label{eq-p2p-guest-subroute}\\
	&kept_i = 1					&i|{\tt propCities}[host][i]=true	\label{eq-p2p-guest-previouslyProposed}\\
	&kept_i\in \{0,1\}				&i\in\{1, \ldots N-1\}			\label{eq-p2p-guest-var}\\
	&p_i\in \Re^+					&1\leq i < N				\label{eq-p2p-guest-varU}\\
	&x_{ij}\in \{0,1\}				&(i,j)\in\{0,1, \ldots N-1\}^2		\label{eq-p2p-guest-var}
\end{eqnarray}
Consequently, Equations \ref{eq-p2p-guest-obj} and \ref{eq-p2p-guest-subroute} are the same as Equations \ref{eq-TSP-obj} and \ref{eq-TSP-subroute}. Equations \ref{eq-p2p-guest-flowOut} and \ref{eq-p2p-guest-flowIn} are similar to Equations \ref{eq-TSP-flowOut} and \ref{eq-TSP-flowIn} except for the city $i$ to be proposed which has $kept_i=0$. Equation \ref{eq-p2p-guest-nb-of-cities-kept} checks that exactly one city will be proposed to the host.
Equation \ref{eq-p2p-guest-previouslyProposed} ensures that a city previously proposed and returned by the guest will not be proposed again.\footnote{When Equations \ref{eq-p2p-guest-obj}-\ref{eq-p2p-guest-var} are solved in Mechanism \sf{Auction}, the previous sentence reads: ``Equation \ref{eq-p2p-guest-previouslyProposed} ensures that a city previously proposed and returned by the auctioneer will not be proposed again''.}
If $propCities$ indicates that all cities have already been proposed, then the considered guest proposes {\tt null}, which fires transition {\tt\ P2P5\_host\_receives\_negotiation\_rejection}.

Otherwise, the city proposed by the guest fires transition {\tt P2P5\_host\_re\-ceives\_city\_proposed\_by\_guest} and state {\tt P2P6\_host\_proposes\_city\_\-to\_guest} finds the city to be proposed by the host to the guest. This is the second call to CPLEX in this round of interaction. Again, CPLEX solves the above modified \ac{TSP} in Equations \ref{eq-p2p-guest-obj}-\ref{eq-p2p-guest-var} (with a small modification: {\tt host} needs to be replaced by {\tt guest} in Equation~\ref{eq-p2p-guest-previouslyProposed}). The guest may not find a city reducing his route length and, hence, send a {\tt null} reply to the guest. Otherwise, he memorises in $propCities$ not to keep proposing the city he has just proposed.

{\tt P2P7\_guest\_receives\_proposition\_and\_sends\_reply} receives this proposed city. If the city proposed by the host is {\tt null}, then the guest sends a message to confirm the failure of this round of negotiation. Otherwise, he uses a CPLEX to solve the traditional \ac{TSP} in Equations \ref{eq-TSP-obj}-\ref{eq-TSP-var} with all his cities, minus the one previously proposed to the host, plus the one which has just been proposed by the host. If the proposed exchange reduces his route length, then the guest sends an acceptation message and updates his route length by taking this exchange into account, otherwise he sends a {\tt null}.

Finally, transition {\tt P2P8\_host\_receives\_acceptation\_or\_rejection} receives this reply. If the guest accepts the exchange, the host modifies his $allocatedCities$ then updates his $route$ by solving Equations \ref{eq-TSP-obj}-\ref{eq-TSP-var} and setting $propCities[city][guest]=false$ for all his $city$ in order to propose any city to this guest again.

{\tt P2P9\_host\_does\_nothing} and {\tt P2P10\_host\_waits} make the host wait for one millisecond. This tiny pause is requested by AnyLogic in order to allow other salesmen to become host. Otherwise, Salesman 0 would keep the control in all subsequent rounds until he has proposed all his cities, next Salesman 1 would become host as long as he has not proposed all his cities, then Salesman 2, etc.

Finally, note that \sf{P2P} does not loop forever because State {\tt P2P2\_host\\\_invites\_guest} does not send an invitation when $propCities[k][i]=true$ for all Salesmen $k$ and all Cities~$i$, which eventually causes every salesman except one to stop acting as host.

%
%
\subsubsection{\sf{CNP}}

Like \citep{frey03}, our Mechanism \sf{CNP} is inspired by the Contract Net Protocol. \sf{CNP} is described by state charts {\tt CNP\_host} and {\tt CNP\_guest} in Figure~\ref{xfig/salesman.eps}. Similarly to \sf{P2P}, the states and transitions have a name starting by {\tt CNPn\_action} where {\tt n} is the order of activation of the considered element and {\tt action} describes it. \sf{CNP} runs by rounds in which a salesman (host) plays the role of an auctioneer who broadcasts a city to give and the other salesmen (guests) reply by proposing a city to be exchanged. Conversely to \sf{P2P}, \sf{CNP} has several guests.

The detail of this mechanism is as follows.
{\tt CNP0\_initialisation} sets all the entries in $propCities[k][i]$ to $false$; This is performed by all salesmen because the first state in all state charts is always executed.
All salesmen wait in state {\tt CNP1\_guest\_waits\_for\_RFP}.
Transition {\tt CNP2\_host\-\_starts} in all salesmen may fire because its condition only checks that Mechanism \sf{CNP} is selected; This condition fires first in one of the salesmen who becomes the host in this round. (Like \sf{P2P}, \sf{CNP} could have several hosts and we prevent this by allowing only one thread in AnyLogic, as detailed in Section~\ref{duration}.)
The host does the first use of CPLEX in this round to select a city to give by solving the modified \ac{TSP} in Equations \ref{eq-p2p-guest-obj}-\ref{eq-p2p-guest-var}; This city is sent in a Request For Proposals to all the guests in {\tt CNP3\_host\_broadcasts\_RFPs}.
Every guest also uses CPLEX to solve the problem in Equations \ref{eq-p2p-guest-obj}-\ref{eq-p2p-guest-var} in order to make a proposal in {\tt CNP5\_guest\_sends\_proposal}.
When transitions {\tt CNP6\_host\_received\-\_proposal} and {\tt CNP6b\_host\-\_received\-\_pro\-posal} have received all these proposals, the host selects the winner by calling CPLEX to test each proposed city in {\tt CNP9\_host\_sends\_alloca\-tion\_replies}. More precisely, for each city submitted by the guests, the host solves the traditional \ac{TSP} in Equations \ref{eq-TSP-obj}-\ref{eq-TSP-var} with his allocated cities minus the city broadcast in the \ac{RFP} plus the city submitted by the considered guest.
The guest sends an acceptation message to the guest who proposed the city reducing the most his route length. In this case, the guest updates his variables {\tt allocatedCities} and {\tt route} ({\tt route} points to the same cities as {\tt allocatedCities} but in the order minimising the route length).
If no city causes such a reduction, then no acceptation is sent. Finally, the guest sends a rejection message to the other guests.
After the acceptation (respectively, rejection) message has been received by {\tt CNP10\_guest\_recei\-ved\_acceptation} (respectively, {\tt CNP10\_guest\_received\_rejection}), another round may start.

After pure \ac{DO}, we now turn our attention to pure \ac{CO}.

%
%
\subsection{\ac{CO} with constraints of \ac{DO}: \sf{OptDecentr}}

As previously said, \sf{OptDecentr} is a \ac{CO} organisation that mimics \ac{DO}.
In other words, \ac{CA} uses CPLEX in order to find the optimal solution of our \ac{MTSP} constrained by 1-1 exchanges. Please first notice that this constraint of 1-1 exchanges of cities makes all salesmen always keep the same number of cities as in their initial endowment. As a result, the decision variable must identify the salesmen in order to ensure that their number of allocated cities equals their number of cities owned in this endowment.
Consequently, \sf{OptDecentr} uses a \ac{MILP} model with the 3-index decision variable $x_{ijk}$ which equals one only if Salesmen $k$ goes from City $i$ to City~$j$:
\begin{eqnarray}
\min	&U^\textrm{mod} =\sum_{k=1}^m u^\textrm{mod}_k								\label{eq-MTSP-sum-obj}\\
s.t.	&u^\textrm{mod}_k = \sum_{i=0}^{n-1} \sum_{j=0, j\neq i}^{n-1} d_{ij} x_{ijk}	&1\leq k\leq m		\label{eq-MTSP-sum-obj2}\\
	&\sum_{j=1}^{n-1}				x_{0jk}=1		&1\leq k\leq m			\label{eq-MTSP-sum-flowOutDepot}\\
	&\sum_{j=0, j\neq i}^{n-1} \sum_{k=1}^m 	x_{ijk}=1		&1\leq i < n			\label{eq-MTSP-sum-flowOut}\\
	&\sum_{i=0, i\neq j}^{n-1} \sum_{k=1}^m		x_{ijk}=1		&1\leq j < n			\label{eq-MTSP-sum-flowIn}\\
	&\sum_{i=1}^{n-1}				x_{i0k}=1		&1\leq k\leq m			\label{eq-MTSP-sum-flowInDepot}\\
	&p_i - p_j + n.x_{ijk} \leq n-1						&1\leq i, j < n, 1\leq k\leq m	\label{eq-MTSP-sum-subroute}\\
	&\sum_{i=0}^{n-1} \sum_{j=0, j\neq i}^{n-1} x_{ijk} = \sum_{j=0}^{n-1} o_{jk}	&1\leq k\leq m		\label{eq-MTSP-wap}\\
	&(\sum_{l=0, l\neq j}^{n-1} x_{jlk} ) - x_{ijk} \geq 0			&0\leq i, j < n, 1\leq k\leq m	\label{eq-MTSP-sum-enter=leave}\\
	&p_{i}\in \Re^+								&0\leq i < n			\label{eq-MTSP-sum-varU}\\
	&x_{ijk}\in \{0,1\}							&0\leq i,j < n, 1\leq k\leq m	\label{eq-MTSP-sum-var}
\end{eqnarray}
In this model, Equations \ref{eq-MTSP-sum-obj} and \ref{eq-MTSP-sum-obj2} minimise the total distance travelled by the community of salesmen. Equation \ref{eq-MTSP-sum-flowOutDepot} (respectively, \ref{eq-MTSP-sum-flowInDepot}) checks that all salesmen leave (respectively, enter) the depot. Equation \ref{eq-MTSP-sum-flowOut} (respectively, \ref{eq-MTSP-sum-flowIn}) checks that all cities are left (respectively, entered) exactly once. Equation \ref{eq-MTSP-sum-subroute} is the same constraint of sub-route elimination as Equations \ref{eq-ECS-subroute}, \ref{eq-TSP-subroute} and \ref{eq-p2p-guest-subroute}.
Equation \ref{eq-MTSP-sum-enter=leave} ensures that the salesmen entering and leaving a city are the same.
Equation \ref{eq-MTSP-wap} checks that the number of cities allocated to a salesman equals the number of cities he owns in his initial endowment. (The right-hand side in this equation is a constant as Parameter $o_{jk}$ represents the initial endowment (``ownership'') of cities, modelled by $o_{jk}=1$ if City $j$ is ``owned'' by Salesman $k$ at the beginning of the experiment, and $o_{jk}=0$ otherwise.)

%
%
\subsection{\ac{CO}/\ac{DO} hybrids: \sf{Cluster} and \sf{Auction}}

We now introduce our two hybrid organisations, $viz.,$ \sf{Cluster} and \sf{Auction}. They are not pure \ac{DO} since \ac{CA} takes part in the allocation and not pure \ac{CO} because \ac{CA} lets the salesmen locally solve the \ac{TSP} in Equations \ref{eq-TSP-obj}-\ref{eq-TSP-var}. \sf{Cluster} is more coercive because the salesmen are supposed to let \ac{CA} know about all their cities in order to solve the allocation problem, while \sf{Auction} lets them free to never propose some of their cities if they want for some reason ($e.g.,$ a city has an important client they want to keep or not disclose).
Conversely to \sf{P2P} and \sf{CNP} which only perform bilateral exchanges, \sf{Cluster} and \sf{Auction} may involve more than two salesmen per exchange during a round, that is, Salesman $s_1$ may give a city to Salesman $s_2$, $s_2$ give to $s_3, \ldots$, and $s_q$ give to $s_1$ for any $q\leq m$.

\subsubsection{\sf{Cluster}}	
\label{Cluster}
Because of Hypothesis \sf{Hyp.~4} about the use of CPLEX to make all decisions, Mechanism \sf{Cluster} cannot use whichever solving methods from the clustering literature but only those based on a \ac{MILP} formulation. We use the model proposed in \cite[Sec.~5]{rao71}:
\begin{eqnarray}
\min	&D							&						\label{eq-cluster2-obj}\\
s.t.	&D \geq d_{ij}(x_{ik}+x_{jk}-1)				&1\leq i < j<n, 0\leq k < m			\label{eq-cluster2-diameter}\\
	&\sum_{k=1}^{m} x_{ik} = 1				&1\leq i < n					\label{eq-cluster2-alloc}\\
	&1+\sum_{j=1}^{n-1} x_{jk} = \sum_{j=1}^{n-1} o_{jk}	&0\leq k < m					\label{eq-clusterSwap}\\
	&x_{ik}\in \{0,1\}, D>0					&1\leq i < n, 0\leq k < m			\label{eq-cluster2-var}
\end{eqnarray}
In this model, decision variable $x_{ik}$ is a binary equal to one only when City $i$ is allocated to Salesman/Cluster~$k$.
Equations \ref{eq-cluster2-obj} and \ref{eq-cluster2-diameter} are the objective function which minimises the diameter of the cluster which has the largest diameter.
More precisely, Equation \ref{eq-cluster2-diameter} includes the linearisation of $D_k \geq d_{ij}x_{ik}x_{jk}$ which ensures that the diameter of Cluster $k$ is at least the maximum distance between any two Cities $i$ and $j$ allocated to this cluster; Please notice that such a formulation defines circular clusters which may hence overlap.
Like in any other allocation problem, Equation \ref{eq-cluster2-alloc} checks that every city is affected to exactly one cluster/salesman.
In addition to the original model by \cite{rao71}, we add Equation \ref{eq-clusterSwap} in order to ensure that the size of the clusters(/salesmen) correspond to the number of cities provided by every (cluster/)salesman, that is, if Salesman $k$ owns $\sum_j o_{jk}$ cities (again, $o_{jk}$ is a constant which equals one only if Salesman $k$ ``owns'' City $j$ in his initial endowment), then a cluster with $\sum_j o_{jk}$ cities must exist in order to be allocated to~$k$. 

As previously said, Mechanism \sf{Cluster} operates in a single round: \ac{CA} first solves the problem in Equations \ref{eq-cluster2-obj}-\ref{eq-cluster2-var} to allocate $N$ cities to every salesman\footnote{Like in Footnote~\ref{fnote}, $N$ may not be that same for all salesmen since it is a local variable.}, then  these salesmen locally solve the traditional \ac{TSP} in Equations~\ref{eq-TSP-obj}-\ref{eq-TSP-var}.

\subsubsection{\sf{Auction}}

\begin{figure}[t]
    \begin{center}
    \centering
    \includegraphics[width=9cm]{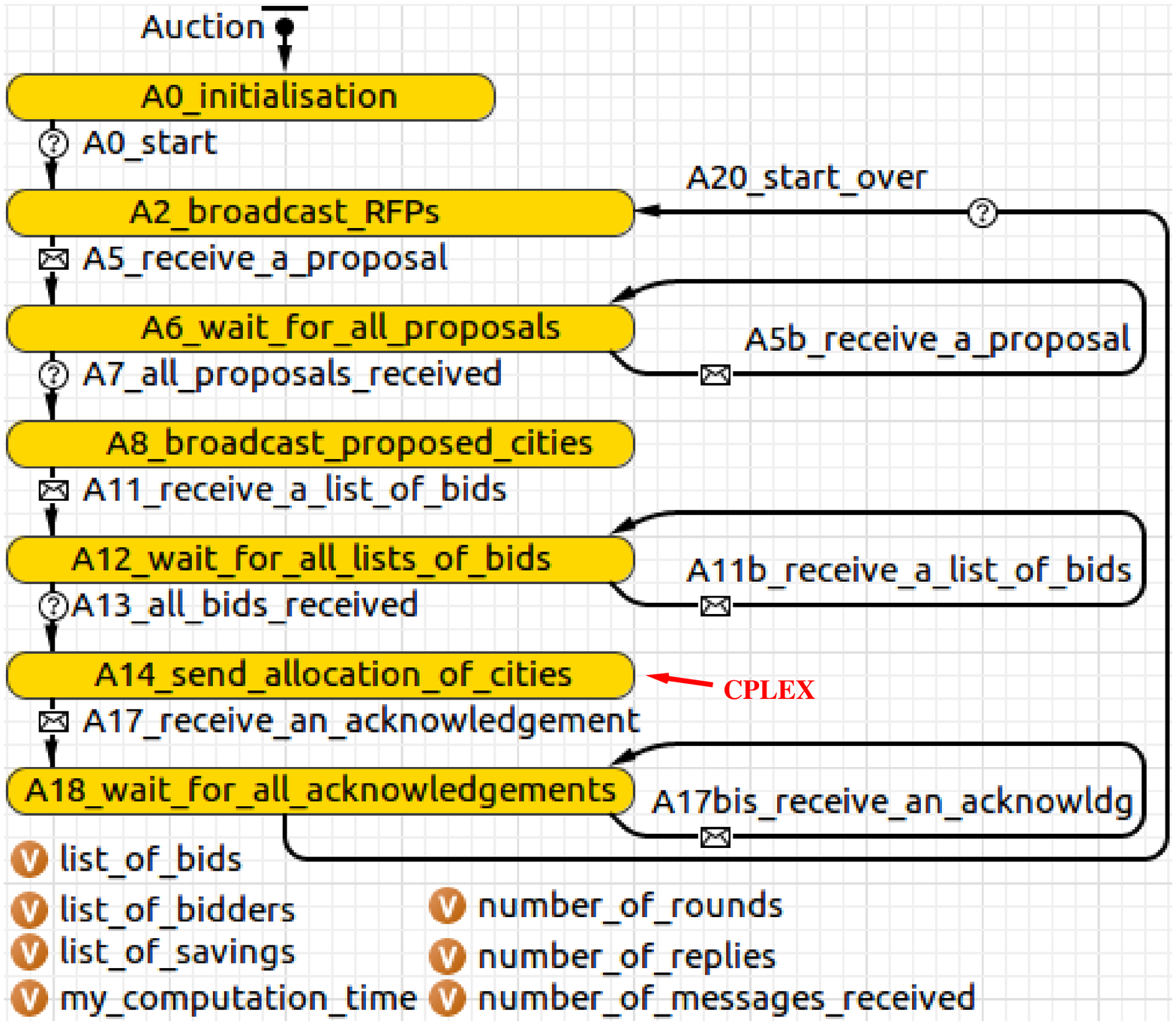}
    \caption{State chart of \sf{Auction} in the \acf{CA}.}
    \label{xfig/centralAuthority.eps}
    \end{center}
\end{figure}

\sf{Auction} is an organisation in which \ac{CA} is an auctioneer who is thus less coercive than in \sf{Cluster}. Conversely to \sf{Cluster}, \sf{Auction} operates in several rounds. In each round, every salesman gives a city $A$ to \ac{CA} who either gives it back if no other salesman wants it, or gives a city $B$ proposed by another salesman if exchanging $A$ and $B$ reduce the length travelled by both salesmen.
Figure \ref{xfig/centralAuthority.eps} and the top right corner of Figure \ref{xfig/salesman.eps} detail the states and transitions in this organisation. As above, their name has the form {\tt A$n$\_action} where {\tt n} helps the reader understand the order of their activation and {\tt action} summarises their goal.

At the beginning of a round, all salesmen wait in {\tt A1\_wait\_for\_RFP} until \ac{CA} sends them a Request For Proposals in {\tt A2\_broadcast\_RFPs}. When a salesman receives this message, he uses CPLEX to look for a city to give in {\tt A4\_propose\_a\_city\_to\_give} by solving the modified \ac{TSP} in Equations \ref{eq-p2p-guest-obj}-\ref{eq-p2p-guest-var}. When all salesmen have replied by sending either a city or {\tt null} to \ac{CA}, \ac{CA} broadcasts this list of replies to all salesmen in {\tt A8\_broadcast\_proposed\_cities}. For each city in this list, every salesman uses CPLEX to compute a bid for it in {\tt A10\_bid\_on\_every\_proposed\-\_city}. A bid for a city is the additional distance travelled to visit it, or, more technically, as the differences between:
	\begin{itemize}
		\item the shortest length of the route visiting their remaining $N-1$ cities, that is, their allocated cities except the one proposed to the auctioneer in {\tt A4\_propose\_a\_city\_to\_give}. This is obtained by each salesman by solving the \ac{TSP} in Equations~\ref{eq-TSP-obj}-\ref{eq-TSP-var} once.
		\item the shortest length of the route visiting their remaining $N-1$ cities plus the city proposed by one of the other salesmen. This is obtained by each salesman by solving the \ac{TSP} in Equations~\ref{eq-TSP-obj}-\ref{eq-TSP-var} for each city in the list of bids.
	\end{itemize}

When all salesmen have returned their list of bids, \ac{CA} uses CPLEX to solve an allocation problem in {\tt A14\_send\_allocation\_of\_cities}. In order to describe this problem, let us call {\tt savings}$[k][i]$ the bid of Salesman $k$ for City $i$ and binary decision variable $x_{ki}$ equals $1$ only if City $i$ is allocated to Salesman~$k$. For simplicity, we write $m$ the number of salesmen who have not left the auction before the current round.
The allocation model solved by \ac{CA} is described by Equations \ref{eq-auction-allocation-obj}-\ref{eq-auction-allocation-var}. The objective in Equation \ref{eq-auction-allocation-obj} allocates the cities such that the total route length of all salesmen is minimum. The constraint in Equation \ref{eq-auction-allocation-individual-selfishness} ensures that, in every auction round, every salesman does not increase his individual route length.
\begin{eqnarray}
\min	&\sum_{k=1}^{m} \sum_{i=0}^{m} {\tt savings}[k][i].x_{ki}						\label{eq-auction-allocation-obj}\\
s.t.	&\sum_{i=0}^{m} x_{ki}=1				&0\leq k < m					\label{eq-auction-allocation-balance1}\\
	&\sum_{k=1}^{m} x_{ki}\leq 1				&0\leq i < m					\label{eq-auction-allocation-balance2}\\
	&\sum_{i=0}^{m} {\tt savings}[k][i].x_{ki} \leq {\tt savings}[k][k]	&0\leq k < m			\label{eq-auction-allocation-individual-selfishness}\\
	&x_{ki}\in \{0,1\}					&0\leq i,j\leq m				\label{eq-auction-allocation-var}
\end{eqnarray}


\section{Real time spans deduced from sequential simulations}
\label{duration}

After the description of the compared mechanisms, we now present how the computation time span of every organisation is deduced from sequential experiments, that is, experiments running on a single thread of AnyLogic. In fact, our experimentation involves up to 10 agents (9 salesmen + 1 \ac{CA}) while each of our computers has a CPU with 4 cores only. Thus, configuring AnyLogic to simulate this parallelism would require studying how AnyLogic schedules up to 4 agents in parallel, then deduce how 10 agents would have behaved in reality.
Instead, we prefer to make AnyLogic carry out all computations sequentially, then infer how (pure and mixed) \ac{DO} would occur concurrently in real life.
The time span of the CPLEX computation (duration between the beginning of the first computation and the end of the last one) in an experiment is deduced as follows.
\begin{itemize}
	\item \sf{OptDecentr} and \sf{FullCentr}: \ac{CO} uses no parallelism and the computation time span is thus equal to the computation time recorded in our sequential experiments. More technically, this duration is the difference between the two {\tt System.currentTimeMillis()} after and before {\tt cplex.solve()}. This is also the Java code to measure the duration of all CPLEX calls in all our organisations below.
	\item \sf{Auction}: Figure~\ref{fig-computationTime-auction}a illustrates how AnyLogic sequentially runs the operation of two salesmen, and Figure~\ref{fig-computationTime-auction}b how reality would look like with parallelism:
	\begin{figure}
		\begin{center}
			\includegraphics[width=14.5cm]{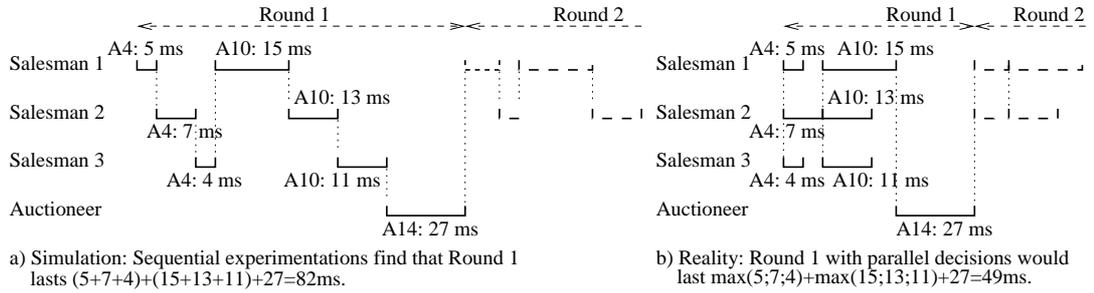}
			\caption{Difference between the computation time (a) in our experiments and (b) in real (b) of one round of \sf{Auction}. (``A$x$:'' is the beginning of the the name in the state charts in Figures \ref{xfig/salesman.eps} and \ref{xfig/centralAuthority.eps}).}
			\label{fig-computationTime-auction}
		\end{center}
	\end{figure}
	\begin{itemize}
		\item \emph{AnyLogic}: Figure~\ref{fig-computationTime-auction}a shows that Salesman 1 calls CPLEX for 5 ms in State {\tt A4\_propose\_a\_city\_to\_give} (summarised as ``A4: 5ms'' in Figures~\ref{fig-computationTime-auction}a and~\ref{fig-computationTime-auction}b), and Salesman 2 and 3 spend 7 ms and 4 ms in this state. Next \ac{CA} (auctioneer) receives the result of these computations and broadcasts it to all salesmen (not shown in Figures~\ref{fig-computationTime-auction} because performed without CPLEX). Later, Salesman 1 computes for 15 ms, Salesman 2 for 13 ms and Salesman 3 for 11 ms in {\tt A10\_bid\_on\_every\_proposed\_city}. Finally, the auctioneer computes for 27 ms in {\tt A14\_send\_allo\-cation\_of\_cities}. The duration observed in AnyLogic is the sum of these durations, as shown in Figure~\ref{fig-computationTime-auction}a.

		\item \emph{Reality}: In real life, all states with the same name can be computed in parallel, as shown in Figure~\ref{fig-computationTime-auction}b.
		It follows that the duration of the CPLEX computations in a state is the maximum of the durations of all salesmen in this state, $i.e.,$ this round would take $\max(5;7;4)+\max(15;13;11)+\max(27)=49$ms.
	\end{itemize}

	\item \sf{Cluster}: The method for \sf{Cluster} is the same as for \sf{Auction} except that (i) there is a single round, and (ii) every salesman solves the \ac{TSP} only once in this round. Shortly, the real-life duration is (i) the time spent by \ac{CA} solving the clustering problem in Equations \ref{eq-cluster2-obj}-\ref{eq-cluster2-var} plus (ii) the maximum of the time spent by the salesmen solving the \ac{TSP} in Equations \ref{eq-TSP-obj}-\ref{eq-TSP-var}.

	\item \sf{NoRealloc}: The method for \sf{NoRealloc} is the same as \sf{Cluster} without \ac{CA}, that is, the real-life duration is the maximum of the time spent by the salesmen solving the \ac{TSP} in Equations \ref{eq-TSP-obj}-\ref{eq-TSP-var}.

	\item \sf{P2P} and \sf{CNP}: The inference of the duration of interactions in pure \ac{DO} is the most complicated because several interactions of various durations may take place concurrently. In contrast, the \ac{CA} in \sf{Auction} ensures that a new round starts only after the end of the previous one. On the contrary, no \ac{CA} synchronises interactions in \sf{P2P} and \sf{CNP} because the state charts in Figure~\ref{xfig/salesman.eps} allow every salesman to be host and guest at the same time in two concurrent interactions.
We assume that interactions do not overlap, that is, an interaction is never stopped between its start and end. This complies with our observation of the operation of AnyLogic when only one thread is used. In addition, using more threads would not change the total duration of interactions and would just make it more complicated to observe what we now explain. This explanation is provided for \sf{P2P} because it is the most complex as several bilateral interactions may occur concurrently; \sf{CNP} is slightly simpler because such overlaps are made impossible by the fact that an interaction always involves all salesmen.

In every round of a \sf{P2P} interaction, we make AnyLogic record the (i) identities of the host and guest in this round, (ii) the starting time in AnyLogic of this round, and (iii) its duration ($i.e.$, sum of (1) the CPLEX computation time of the guest in {\tt P2P4\_guest\_proposes\_a\-\_city}, (2) the time of the host in {\tt P2P6\_host\_proposes\_city\_to\-\_guest}, (3) the time of the guest in {\tt P2P7\_guest\_receives\_propo\-sition\-\_and\_sends\_reply} and (4) the time of host in {\tt P2P8\_host\-\_re\-ceives\-\_acceptation\_or\_rejection} -- please notice that this is a summation because everything also happens sequentially in real life).
	Figure~\ref{fig-computationTime-P2P} illustrates how this information is used after the completion of the mechanism with an example of four interactions named I1, I2, I3 and~I4:
	\begin{figure}
		\begin{center}
			\includegraphics[width=14.5cm]{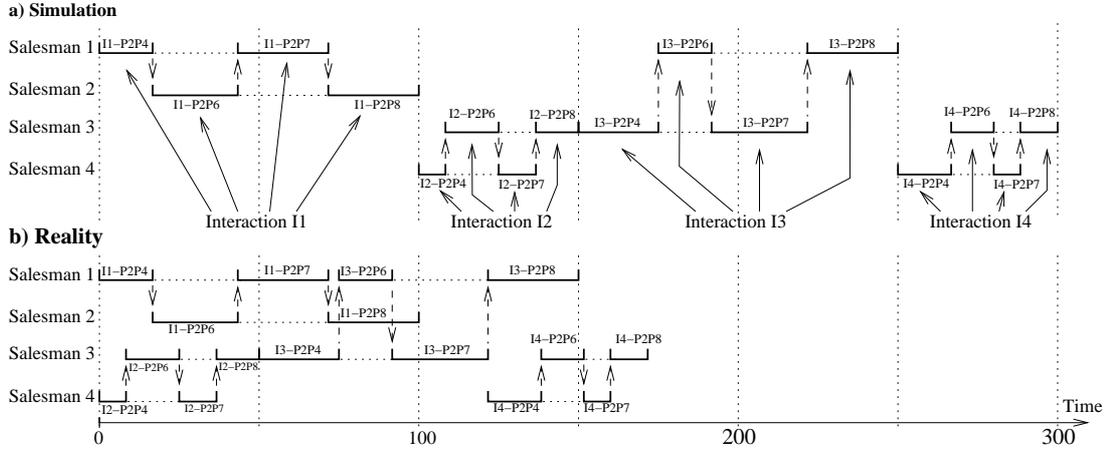}
			\caption{Difference between the computation time (a) in our experiments and (b) in real life in four rounds (called I1, I2, I3 and I4) of \sf{P2P} interactions. (``I$x$'' refers to Interaction I$x$ and ``-P2P$y$'' to the names in the state charts in Figure \ref{xfig/salesman.eps}).}
			\label{fig-computationTime-P2P}
		\end{center}
	\end{figure}
	\begin{itemize}
		\item \emph{AnyLogic}: Salesman 2 is the host of Interaction I1 (Figures~\ref{fig-computationTime-P2P}a and \ref{fig-computationTime-P2P}b only show CPLEX computations, thus {\tt P2P2\_host\_invi\-tes\_guest} is not shown) and the first CPLEX computation is performed by his guest Salesman 2 in State {\tt P2P4\_guest\_propo\-ses\_a\_city}, which is represented by I1-P2P4 in Figure~\ref{fig-computationTime-P2P}a. AnyLogic carries out the computations one after the other: I1-P2P4, then I1-P2P6, I1-P2P7 and I1-P2P8 for Interaction 1, next Interaction 2 with I2-P2P4, I2-P2P6, I2-P2P7 and I2-P2P8, then Interaction 3, etc.

		\item \emph{Reality}: Figure~\ref{fig-computationTime-P2P}b shows that Interactions I1 and I2 will start at the same time in reality since they involve two different pairs of salesmen. Hence, the CPLEX calls in \{I1-P2P4, I1-P2P6, I1-P2P7, I1-P2P8\} would be in parallel with those in \{I2-P2P4, I2-P2P6, I2-P2P7, I2-P2P8\}.
		After that, I3 would start as soon as its host Salesman 1 is available, $i.e.,$ just after I1-P2P7, then this host would wait for the reply of Salesman 2 at the end of I3-P2P3 because this guest would be taking part to I2.
		Similarly, I4 starts just after its host Salesman 3 finished I3-P2P7, which is not shown on Salesman 3 but with the effect on Salesman 4 who computes I4-P2P4.
	\end{itemize}
	Finally, both Figures~\ref{fig-computationTime-P2P}a and \ref{fig-computationTime-P2P}b are incomplete because they show interactions without the initialisation of the salesmen who first solve the \ac{TSP} in Equations~\ref{eq-TSP-obj}-\ref{eq-TSP-var} before $t=0$, and the salesmen in Figures~\ref{fig-computationTime-P2P}b would be ready for their first interaction at different times.

	More technically, the computation time in \sf{P2P} is calculated by updating a vector $clock_k$ after each computation of Salesman~$k$. $clock_k$ represents the total computation time of Salesman $k$ up to the considered interaction. We also call $T_s$ the computation time of CPLEX in state~$s$:
	\begin{itemize}
		\item \emph{Duration of initialisation}: Every salesman $k$ solves the \ac{TSP} in Equations~\ref{eq-TSP-obj}-\ref{eq-TSP-var}, which takes a duration $T_{{\tt P2P0}_k}$. Since these computations are parallel, the duration of this initialisation is $\max_k(T_{{\tt P2P0}_k})$. Hence, at the end of this initialisation, $clock_k=\max_k(T_{{\tt P2P0}_k})$ for all~Salesmen~$k$.
		\item \emph{For every round of bilateral interactions}:
			\begin{itemize}
			\item An interaction starts as soon as its host and guest are both ready, thus an interaction starts at $\max(clock_\textbf{host},clock_\textbf{guest})$. Hence, $clock_\textbf{host} = clock_\textbf{guest} = \max(clock_\textbf{host},clock_\textbf{guest})$.
			\item The end dates of the current round is $clock_\textbf{guest} = clock_\textbf{host} + T_{\tt P2P4\_guest\_proposes\_a\_city} + T_{\tt P2P6\_host\_proposes\_city\_to\_guest} + T_{\tt P2P7\_guest\_receives\_proposition\_and\_sends\_reply}$  and $clock_\textbf{host} = clock_\textbf{guest} + T_{\tt P2P8\_host\_receives\_acceptation\_or\_rejection}$.
			\end{itemize}
		\item \emph{Total computation time}: The searched duration of Mechanism \sf{P2P} is $\max_k(clock_k)$ calculated when all interaction rounds are completed.
	\end{itemize}
\end{itemize}

\section{Numerical experimentation}
\label{experimentation}

Instead of performing Monte Carlo simulations to assess the mechanisms, we allow the reader to replicate our results by generating 130 instances by circular permutations on problem ``CH130'' in TSPLIB.\footnote{\url{http://comopt.ifi.uni-heidelberg.de/software/TSPLIB95/tsp/ch130.tsp.gz}}
In order to describe these circular permutations, let us call $(x_i, y_i)$ the coordinates of the $i^{th}$ city in our simulation and $(X_i, Y_i)$ the coordinates of the $i^{th}$ city in TSPLIB.
For a given number of cities $n$, our Instance zero uses the first $n$ instances in TSPLIB such that City $i$ has $x_i=X_i$ and $y_i=Y_i$ (cities $i\geq n$ in TSPLIB are ignored), $i.e.,$ Salesman 1 is at $(334.5\ldots, 161.7\ldots)$, 2 at $(397.6\ldots, 262.8\ldots)$, 3 at $(503.8\ldots, 172.8\ldots)$, etc. Next, Instance $\Delta Y$ uses $x_i=X_i$ and $y_i=Y_i + \Delta Y$, $i.e.,$ for Instance 1, Salesman 1 is at $(334.5\ldots, 262.8\ldots)$, 2 at $(397.6\ldots, 172.8\ldots)$, 3 at $(503.8\ldots, 384.6\ldots)$, etc.
(We noticed that Instance $\Delta Y=122$ is often very long to solve for several mechanisms.)

This generation of instances allows us to present results in which the mechanisms work on the same instances.
For example, the ratios in Figures \ref{fig-ratios-infinity-OptDecentr}, \ref{fig-ratios-infinity-FullCentr}, \ref{fig-ratios-30minutes-OptDecentr}, \ref{fig-ratios-30minutes-FullCentr} are obtained by (i) setting $n$ and $m$, (ii) comparing all mechanisms ($i.e.,$ computing the ratios of the total route length of two mechanisms) on Instance $\Delta=0$, (ii') repeating ii for $\Delta$ varying between 1 and 129, and (iii) writting in these figures the fifth and ninth deciles of these 130 ratios.

Our numerical experimentation have been performed on the 17 Personal Computers of a student laboratory in the department of Industrial Engineering at INSA-Lyon, Lyon, France.
These computers are all identical and run Windows 7 professionnel SP1 64 bits on Intel Core i5-3470 CPU@3.20GHz with 8.00 Gb RAM.
The softwares used were IBM ILOG CPLEX 12.6.3.0 and AnyLogic~7.3.2.

\subsection{Results with a time limit of 24 hours}

\begin{figure}
	\begin{center}
	\includegraphics[width=6.268cm,height=5cm]{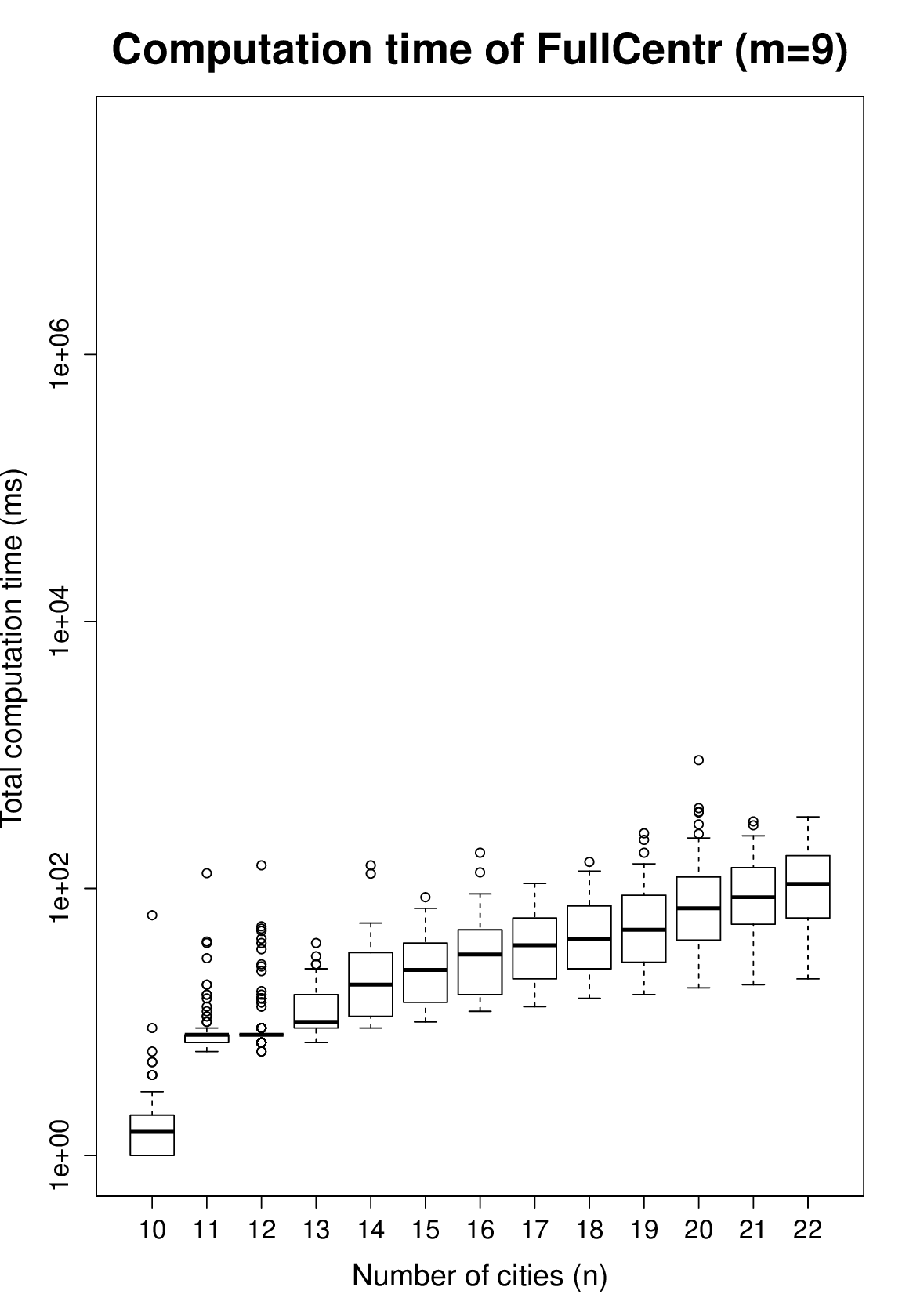}
	\includegraphics[width=6.268cm,height=5cm]{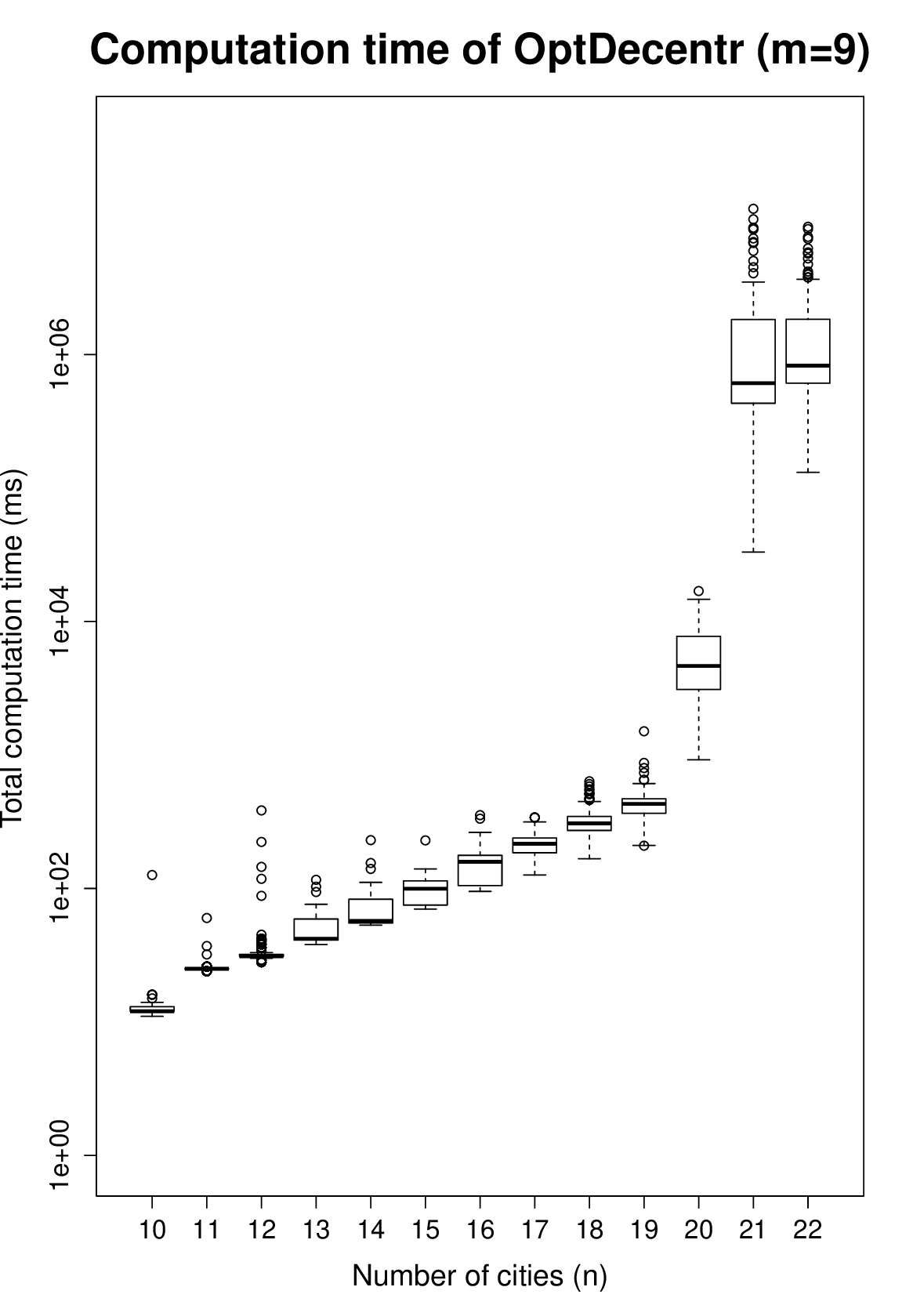}
	\includegraphics[width=6.268cm,height=5cm]{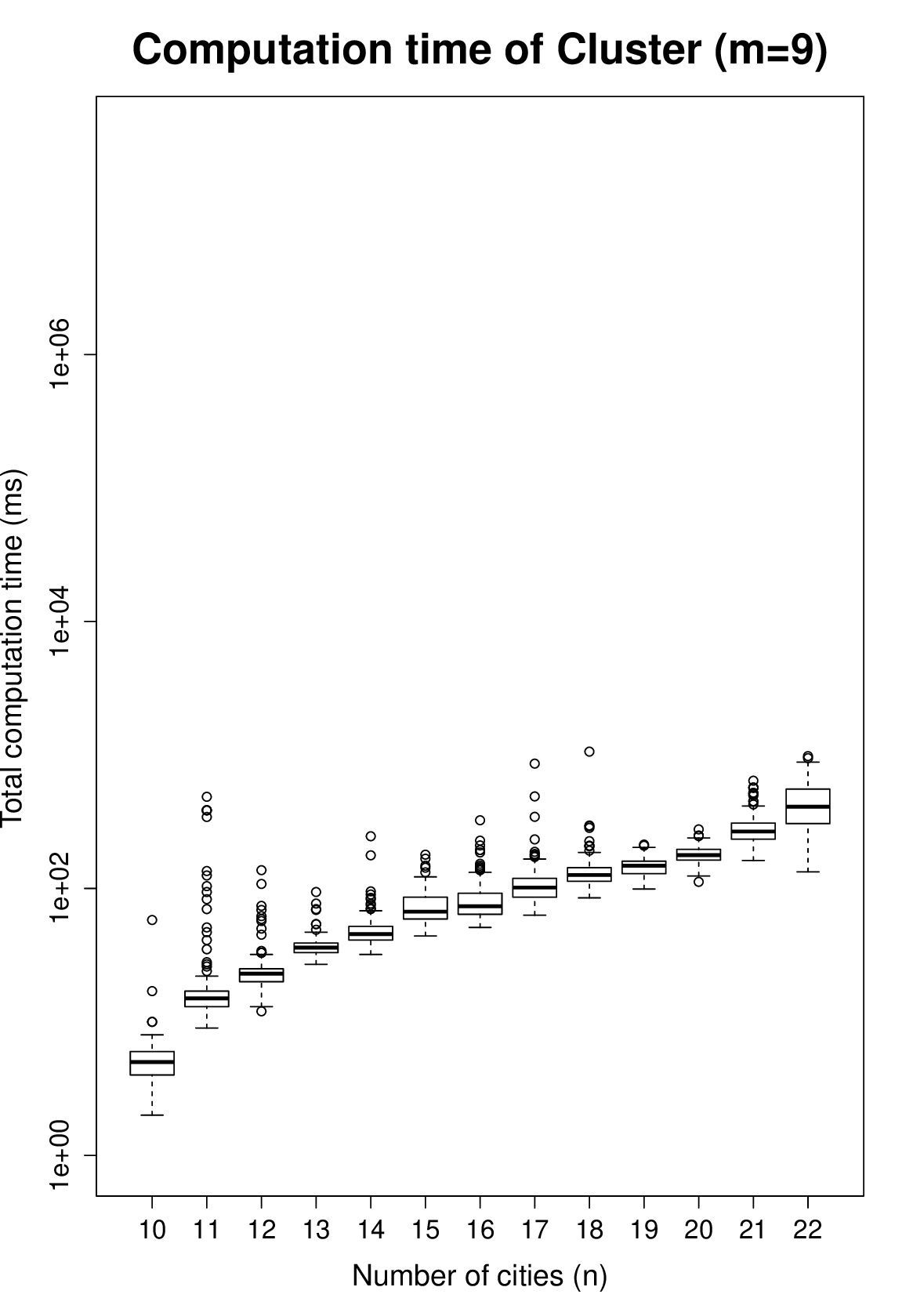}
	\includegraphics[width=6.268cm,height=5cm]{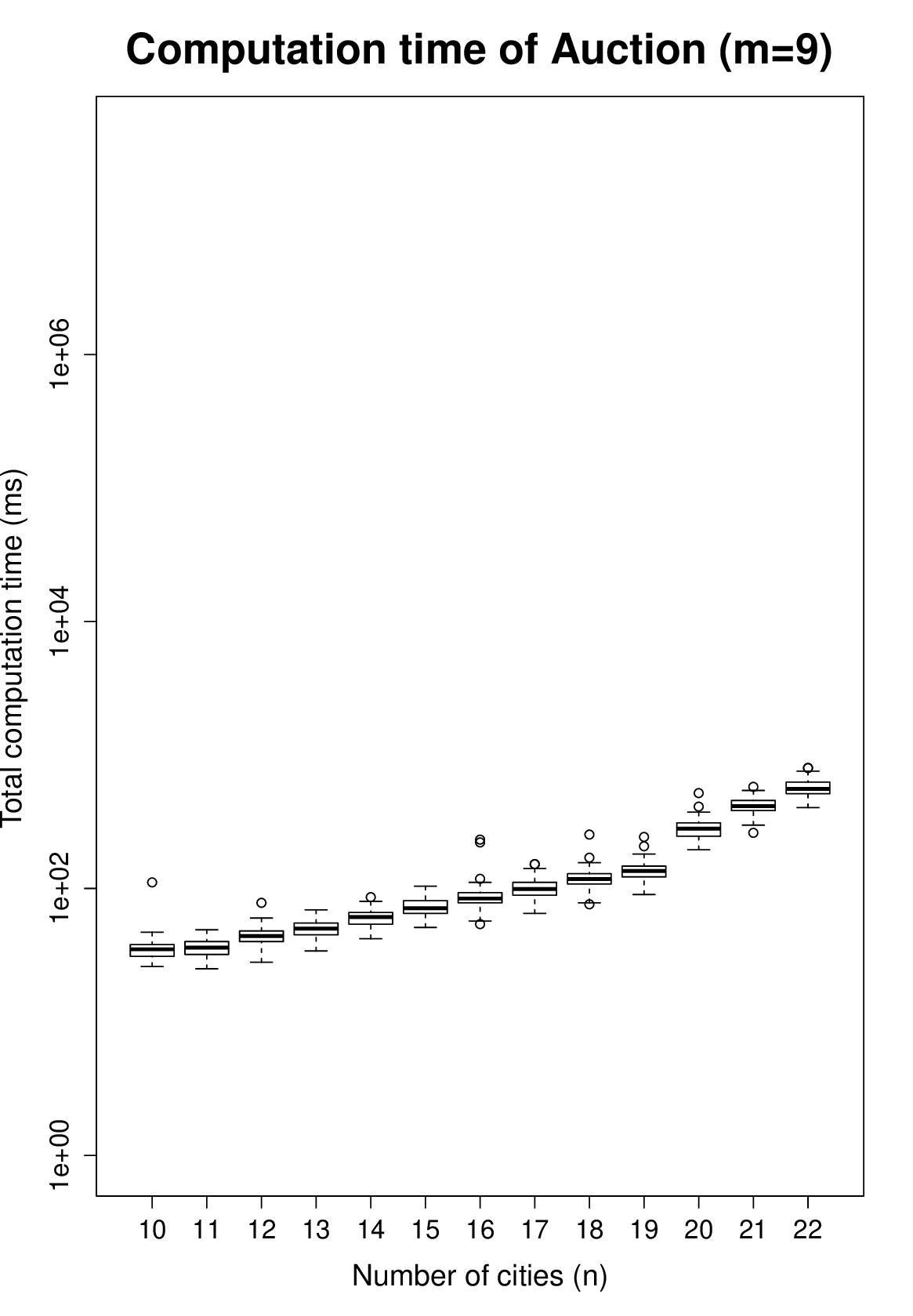}
	\includegraphics[width=6.268cm,height=5cm]{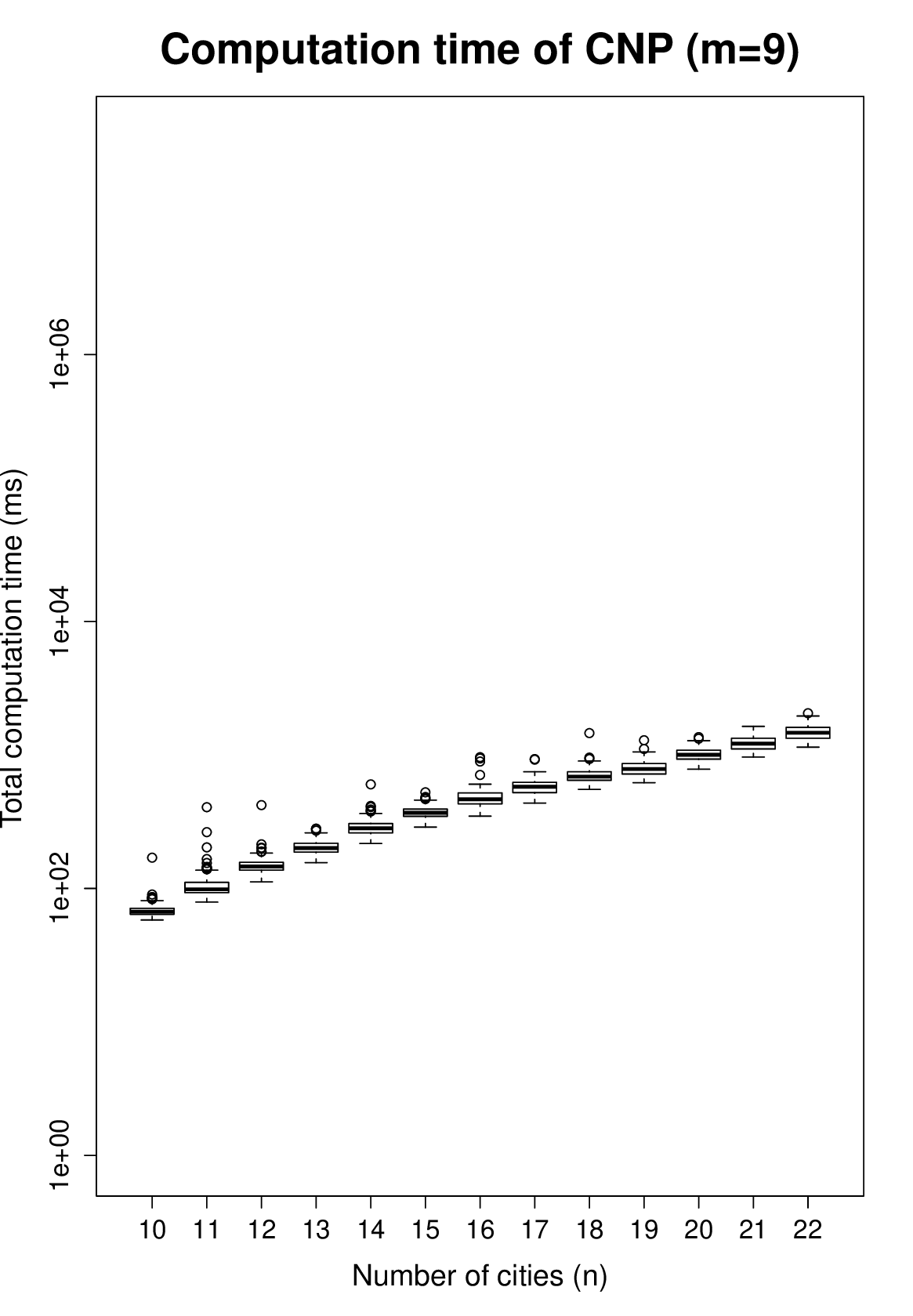}
	\includegraphics[width=6.268cm,height=5cm]{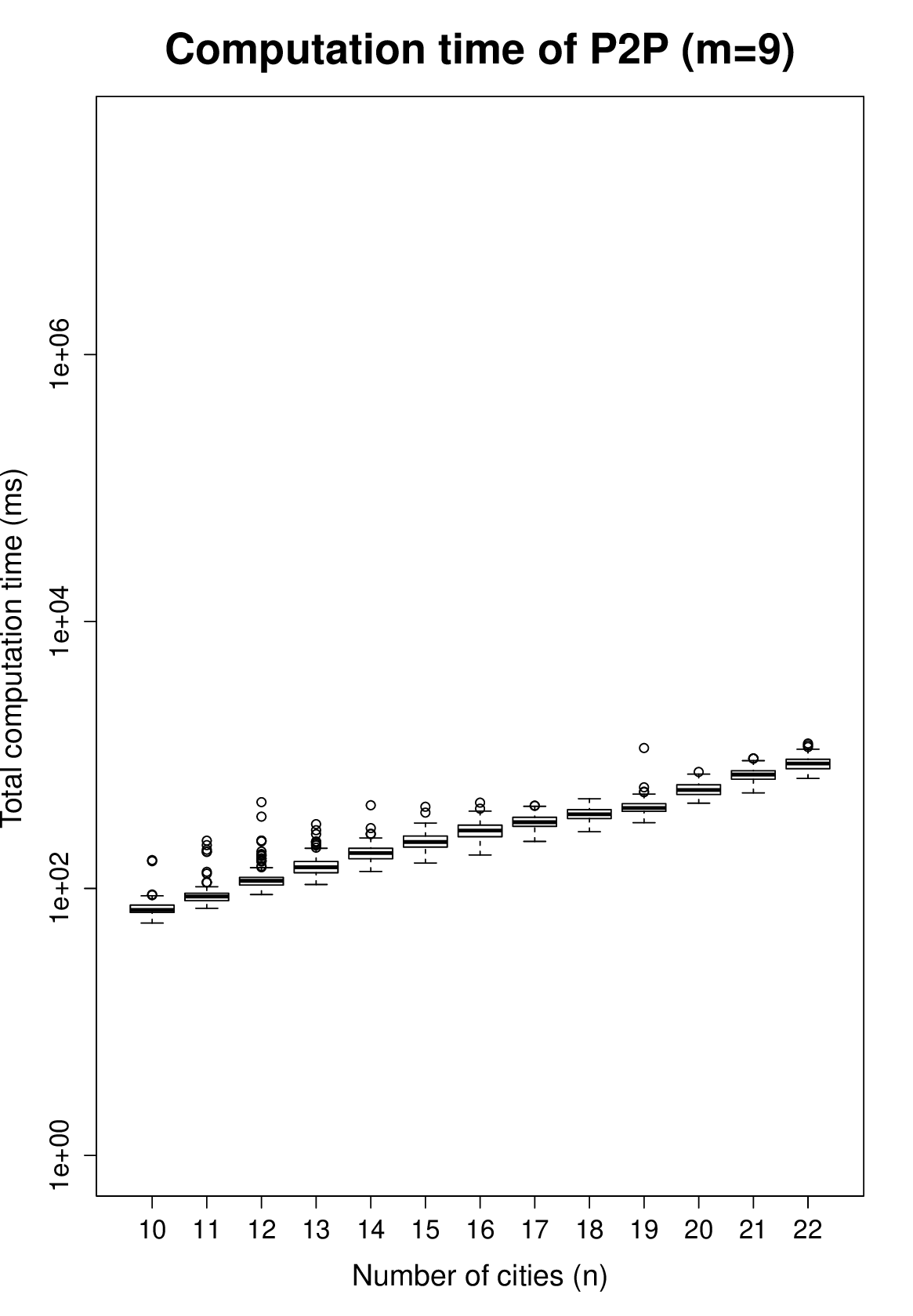}
	\includegraphics[width=6.268cm,height=5cm]{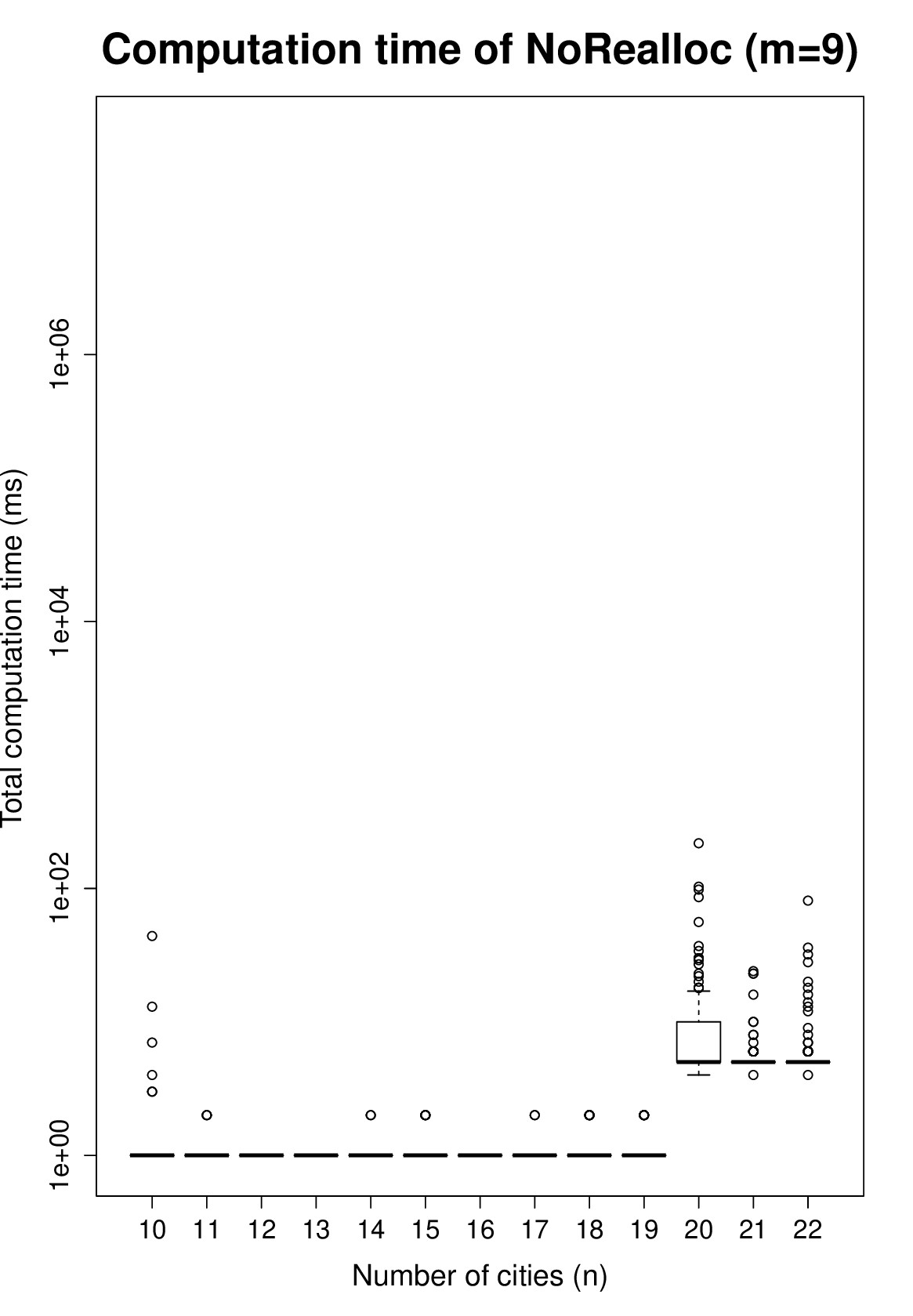}
	\caption{Box-plots of the computation time of the 130 instances for $m=9$ salesmen and various number of cities $n$.}
	\label{fig-boxplots-infinity-TOTcmpT}
	\end{center}
\end{figure}

We consider two metrics to assess our mechanisms, $viz.,$ total route length (also referred to as quality of the solution) and computation time.
In this article, all graphs in a same figure use the same ranges on the y-axis.
For each mechanism, Figure \ref{fig-boxplots-infinity-TOTcmpT} shows the box-plots of the computation time of the 130 instances for various number of cities~$n$ when there are $m=9$ salesmen. 
The main point to notice is the quick increase of the duration of \sf{OptDecentr} from $n=20$, which corresponds $(n-1)/m \approx 2.1$ cities per salesman (``minus one'' prevents counting the depot). 
In this figure, the number of salesmen $m$ is increased until one of the 130 instances cannot reach its end within 24 hours. As can be seen, this limit is set by \sf{OptDecentr} which cannot find the optimal solution of at least one of the 130 instances for $m=9$ salesmen and $n=23$ cities.
We do not show the equivalent of Figure \ref{fig-boxplots-infinity-TOTcmpT} with the quality of the solution because it would be difficult to see a difference between the mechanisms.

\begin{figure}
	\begin{center}
	\includegraphics[width=6.268cm,height=11cm]{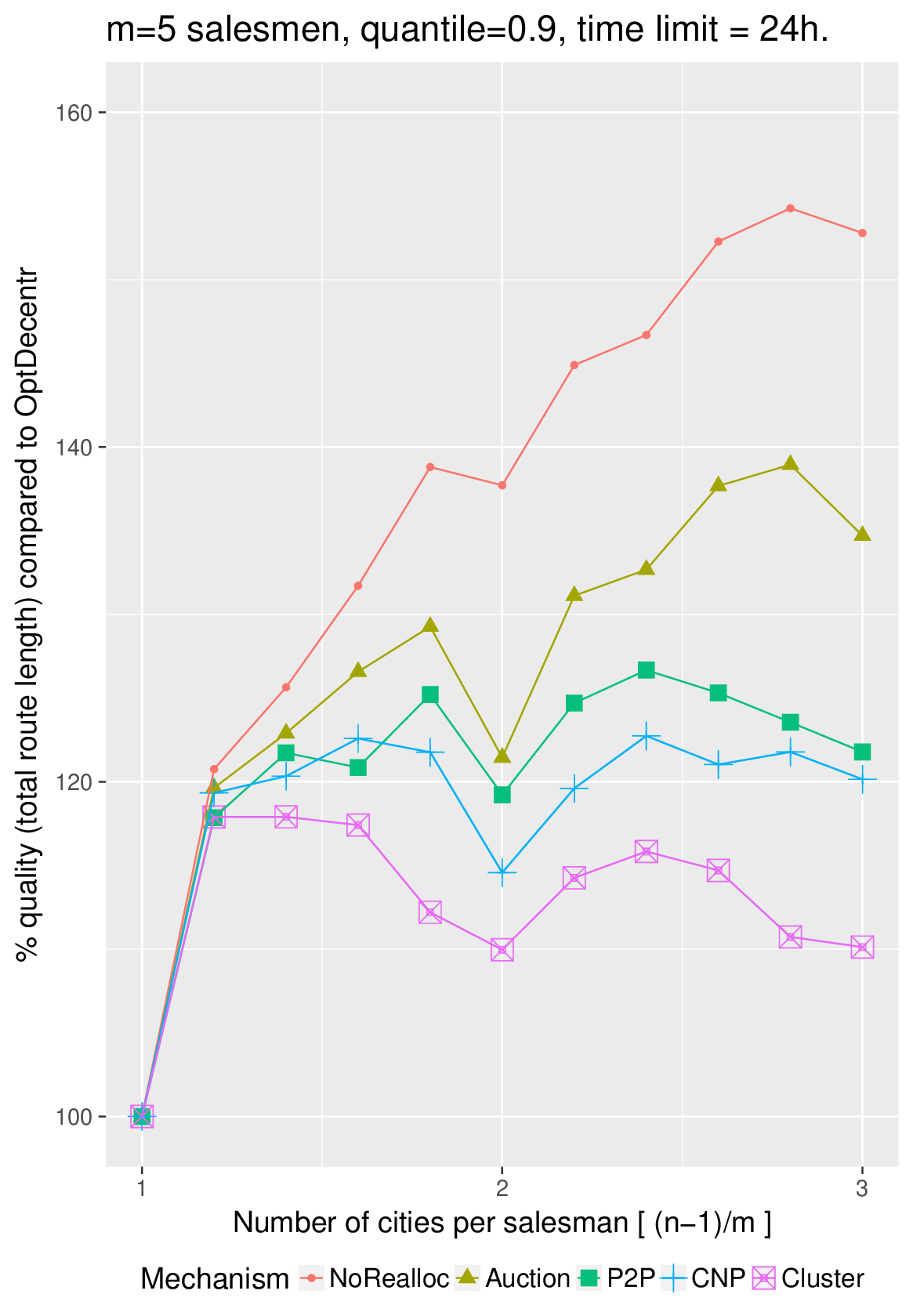}
	\includegraphics[width=6.268cm,height=11cm]{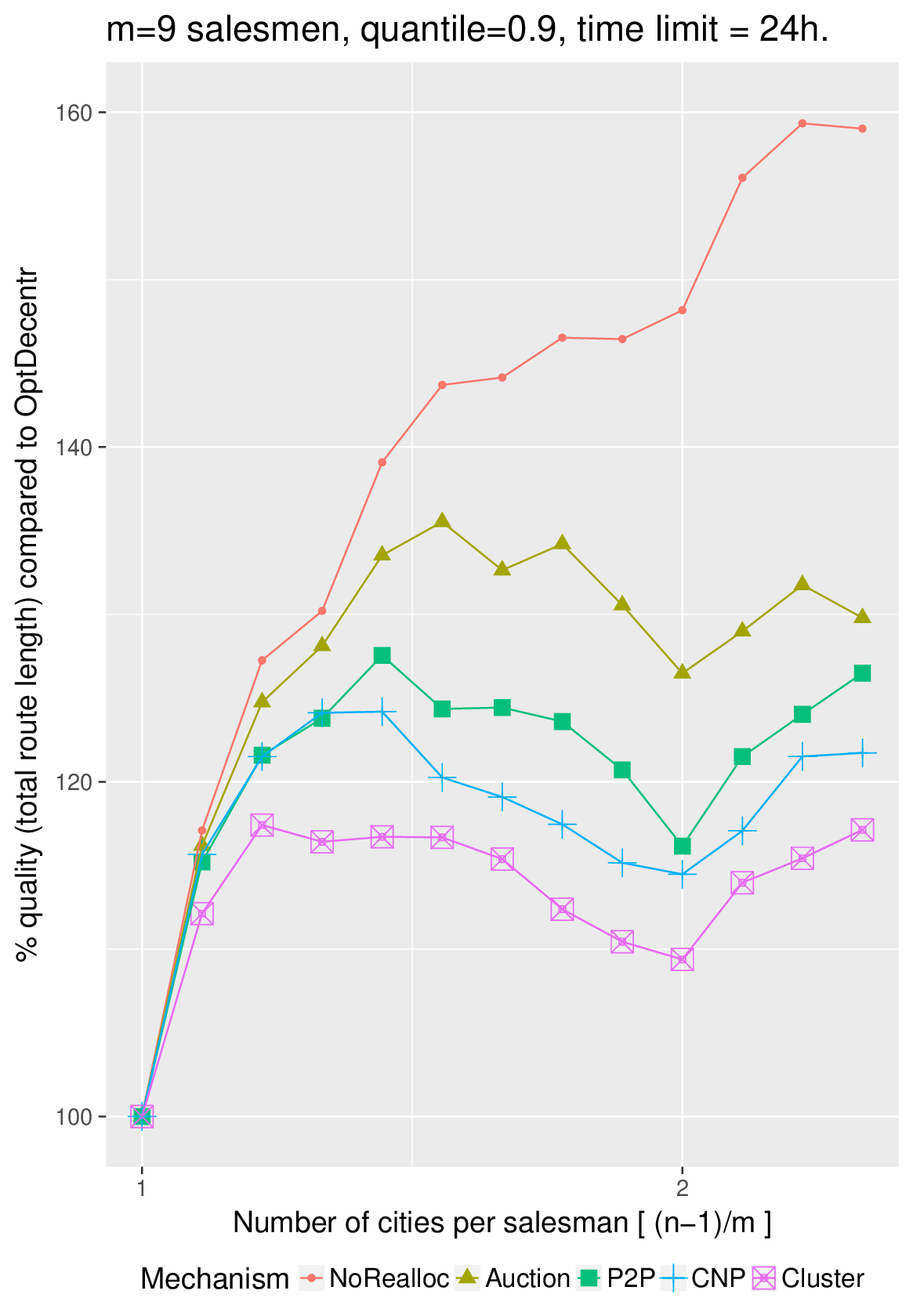}
	\includegraphics[width=6.268cm,height=11cm]{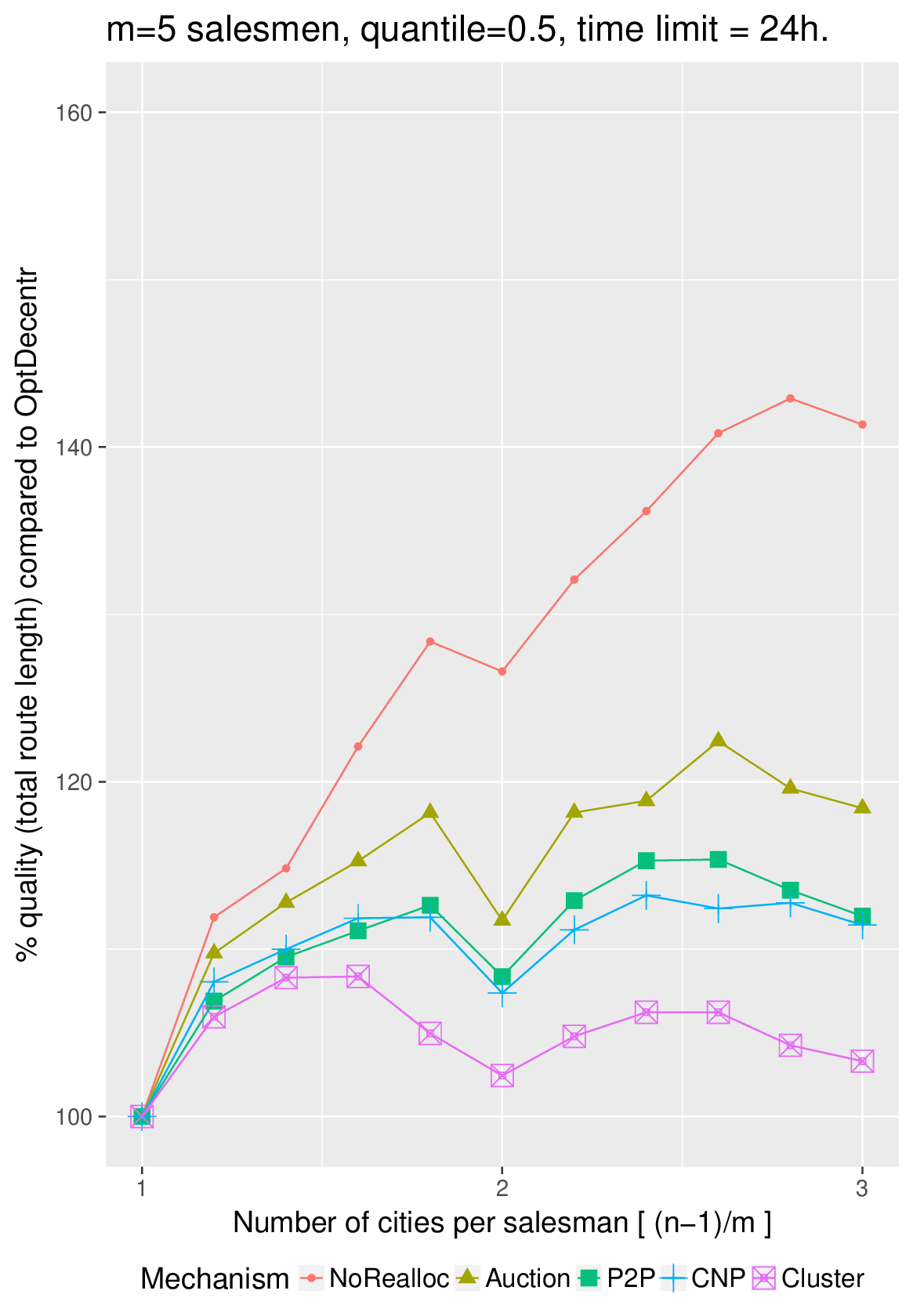}
	\includegraphics[width=6.268cm,height=11cm]{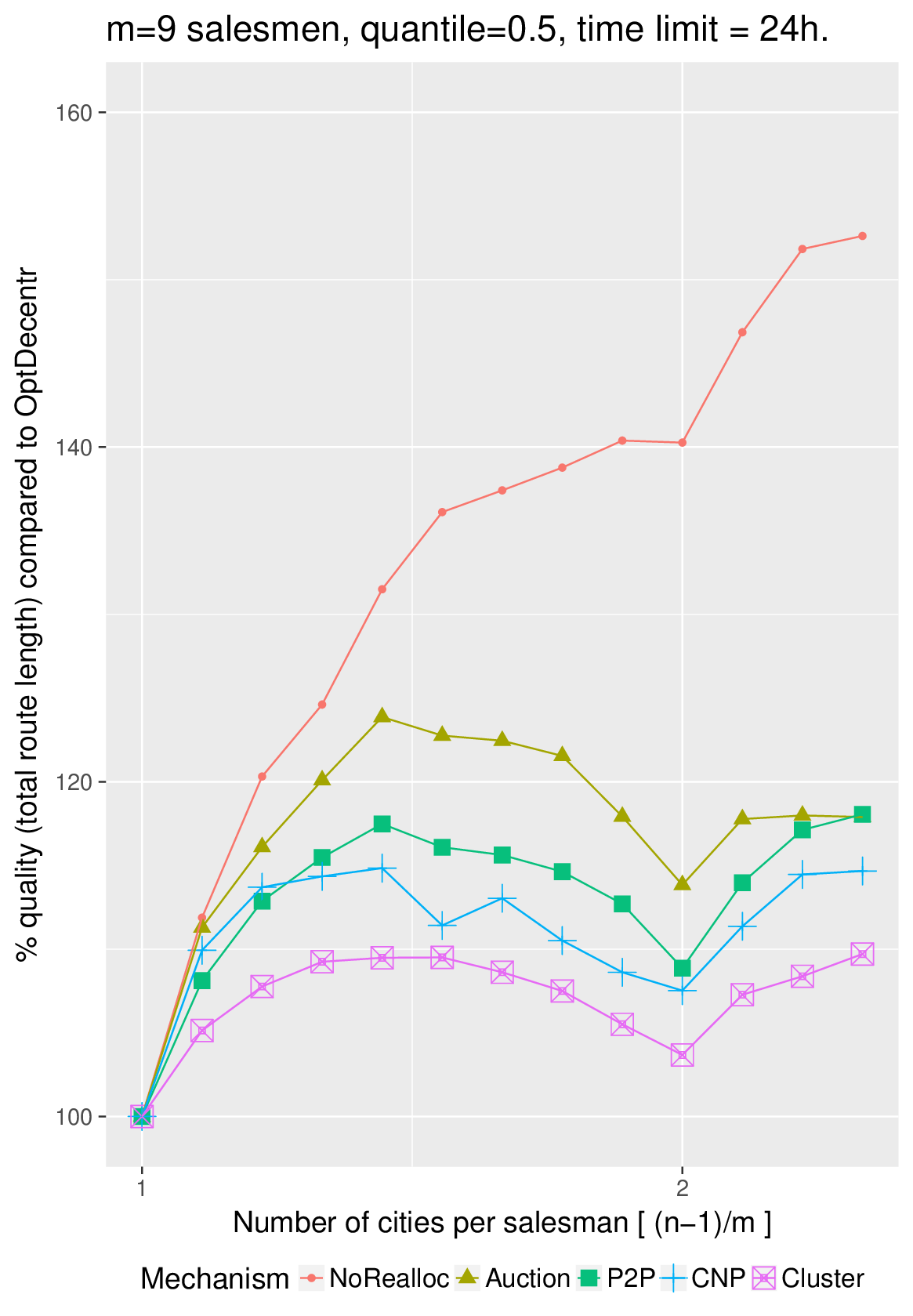}
	\caption{Ratios of quality compared to OptDecentr for $m=5$ (left) and $m=9$ (right) salesmen with no time limit.}
	\label{fig-ratios-infinity-OptDecentr}
	\end{center}
\end{figure}

\begin{figure}
	\begin{center}
	\includegraphics[width=6.268cm,height=11cm]{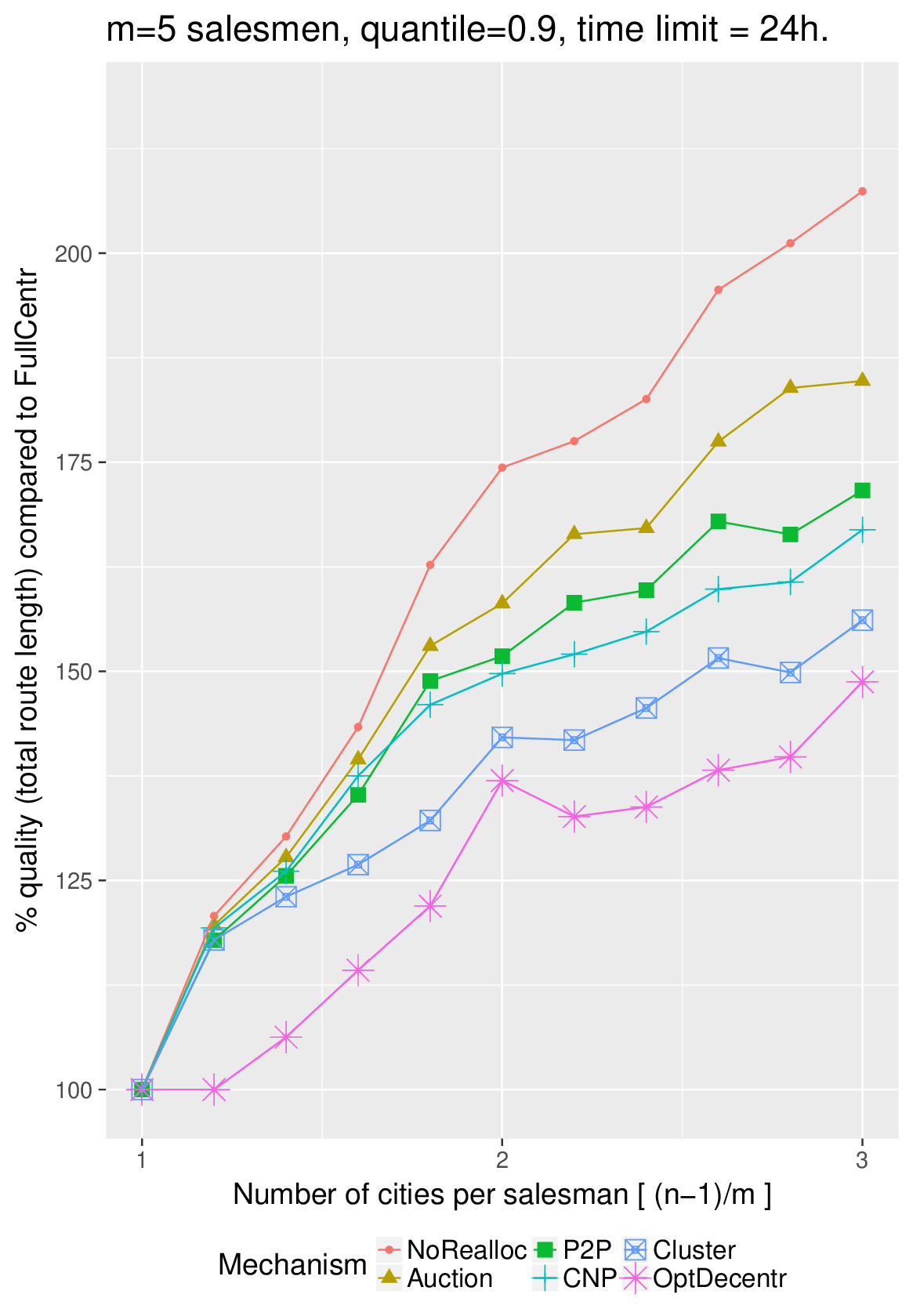}
	\includegraphics[width=6.268cm,height=11cm]{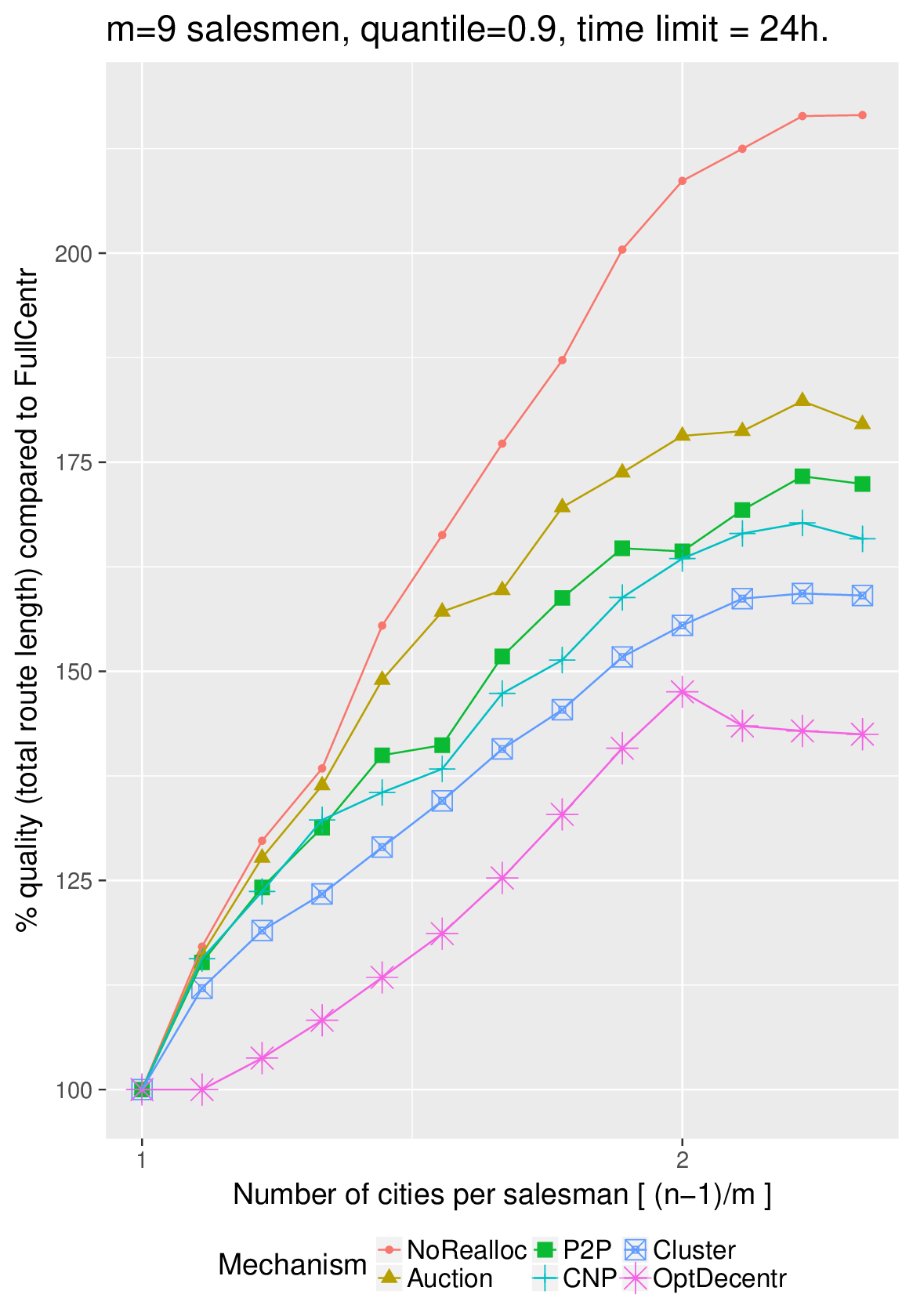}
	\includegraphics[width=6.268cm,height=11cm]{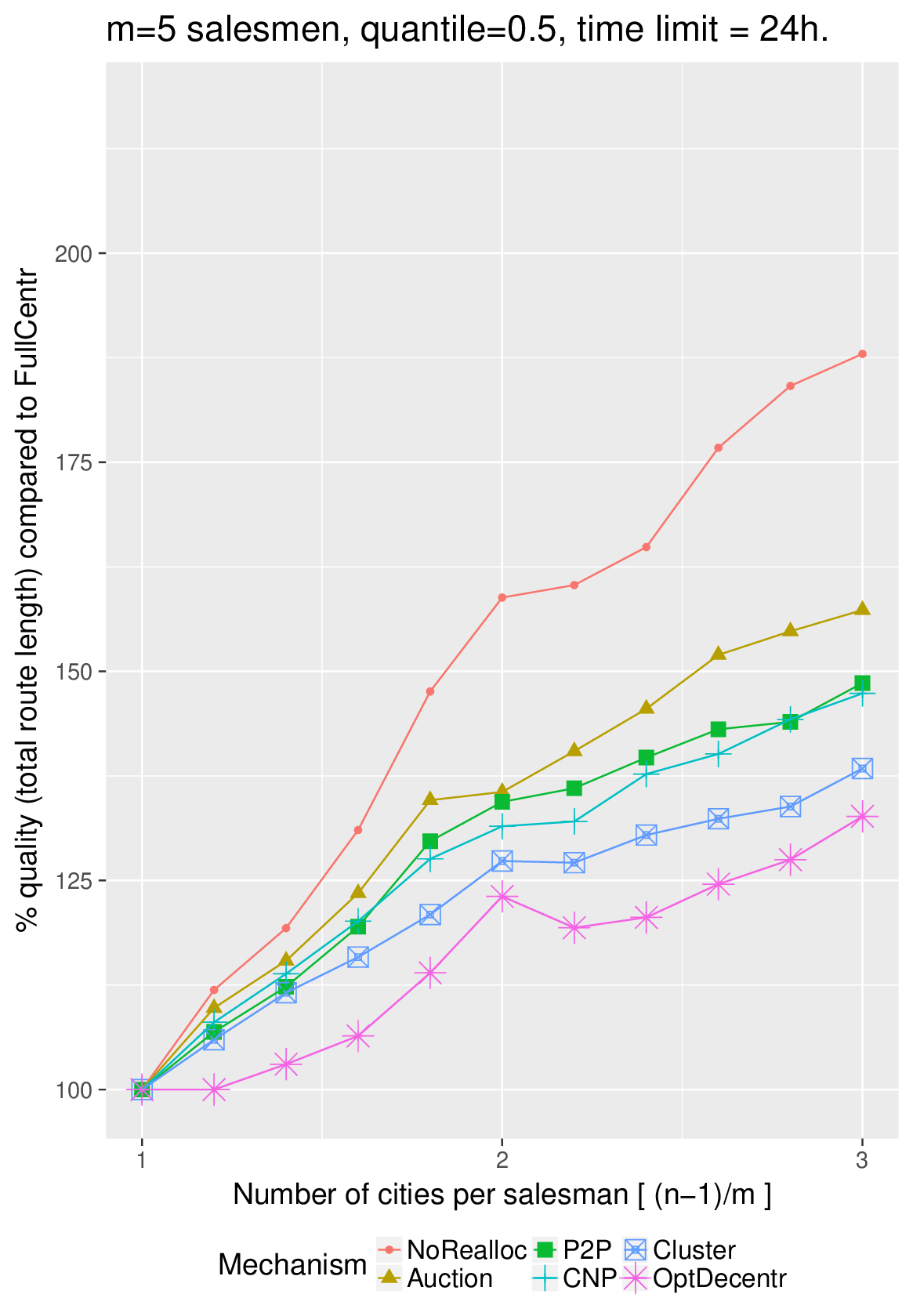}
	\includegraphics[width=6.268cm,height=11cm]{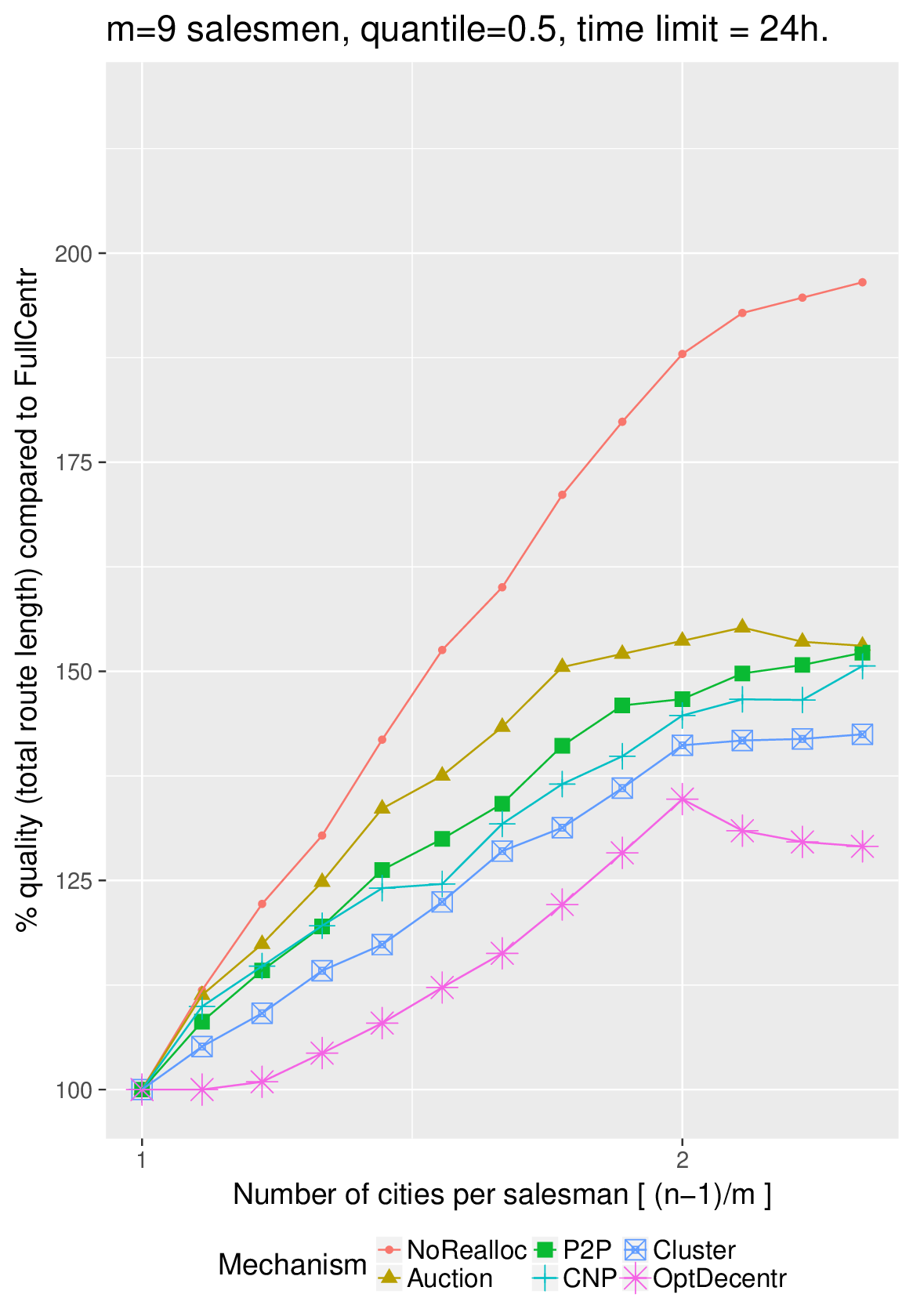}
	\caption{Ratios of quality compared to FullCentr for $m=5$ (left) and $m=9$ (right) salesmen with no time limit.}
	\label{fig-ratios-infinity-FullCentr}
	\end{center}
\end{figure}

Instead, we show Figures \ref{fig-ratios-infinity-OptDecentr} and \ref{fig-ratios-infinity-FullCentr} which show ratios of quality (compared to \sf{OptDecentr} in Figure \ref{fig-ratios-infinity-OptDecentr} and to \sf{FullCentr} in Figure \ref{fig-ratios-infinity-FullCentr}) for $m=5$ (left) and $m=9$ salesmen (right).
More precisely, Figure \ref{fig-ratios-infinity-OptDecentr} compares the total route length found by the mechanisms solving the \ac{MTSP} constrained by 1-1 exchanges and this comparison is carried out against the optimal value found by \sf{OptDecentr}. As said above, for given instance $\Delta Y$ and values of $m$ and $n$, the route length found by a mechanism is divided by the route length found by \sf{OptDecentr} for each of the 130 instances; Both top graphs in \ref{fig-ratios-infinity-OptDecentr} show the ninth decile of these 130 ratios and both bottom graphs show their median.
Figure \ref{fig-ratios-infinity-FullCentr} is computed the same way, but the base of comparison is \sf{FullCentr} instead of \sf{OptDecentr}.
In other words, Figure \ref{fig-ratios-infinity-OptDecentr} compares mechanisms solving the same problem, while Figure \ref{fig-ratios-infinity-FullCentr} compares \ac{DO} mechanisms to the \ac{CO} mechanism.

We think that the most interesting points are the presence of plateaus.
Some mechanisms seem to reach a plateau in Figure \ref{fig-ratios-infinity-OptDecentr}: \sf{Cluster} seems to be about 5\% worse, \sf{CNP} about 11\% worse, \sf{P2P} about 12\% and \sf{Auction} about 20\% worse than \sf{OptDecentr} with the median for both $m=5$ and $m=9$, and the ninth decile also seems to stabilise on slightly higher figures.
This indicates that these decentralised mechanisms do not explore the entire search space of the possible 1-1 exchanges since the \ac{CA} in \ac{DO} is sometimes able to find a better allocation which satisfies the selfishness of the salesmen.
Unfortunately, there does not seem to be such plateaus in Figure \ref{fig-ratios-infinity-FullCentr}, because either they do not exist or $n$ has not been increased enough to reach them. We believe that the latter explanation is true. In order to check that we are right, the next subsection reduces the time limit and keeps increasing $n$ even when the computation time reaches this limit.

\subsection{Results with a time limit of 30 minutes}
\label{result-30minutes}

\begin{figure}
	\begin{center}
	\includegraphics[width=6.268cm,height=10.5cm]{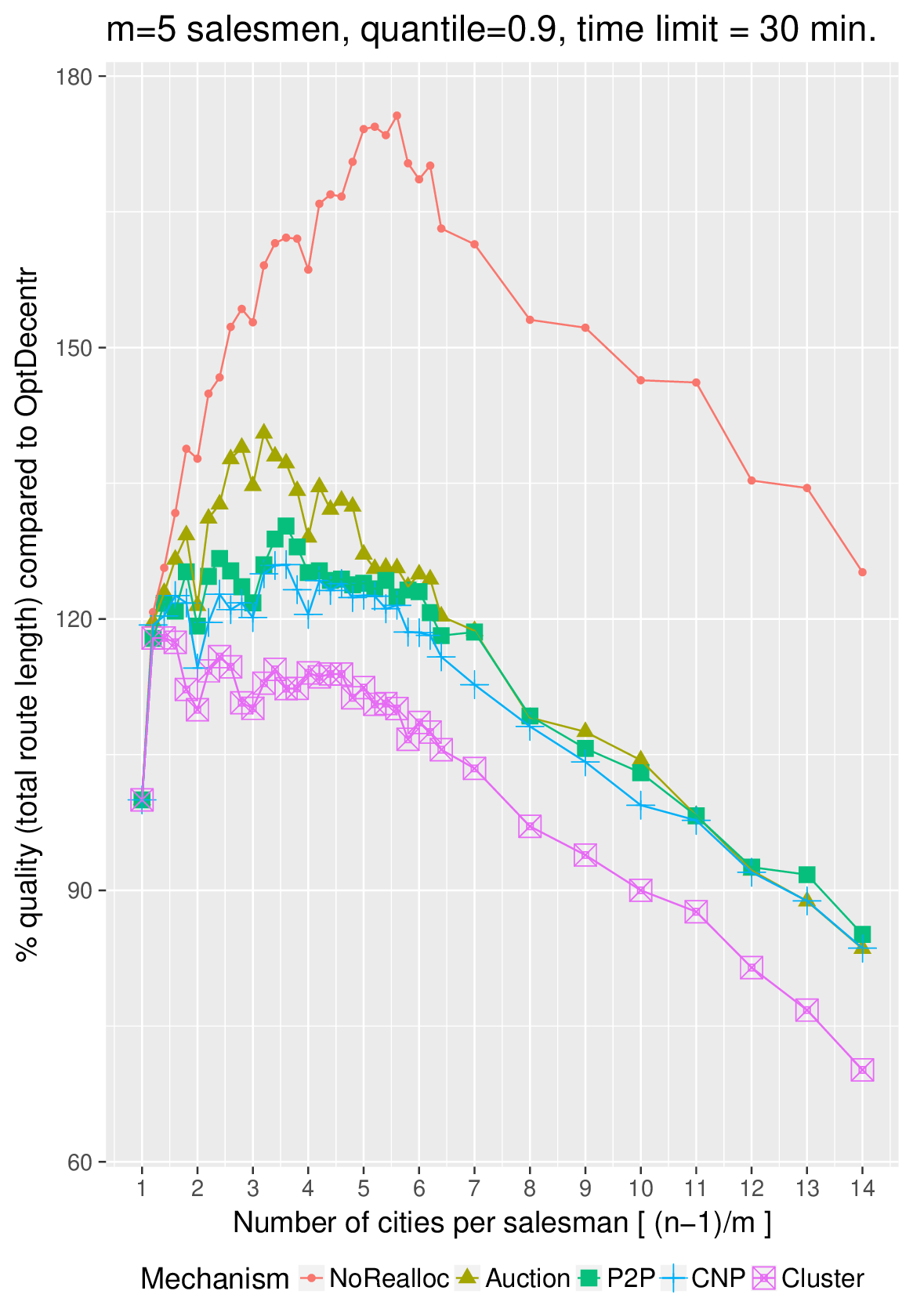}
	\includegraphics[width=6.268cm,height=10.5cm]{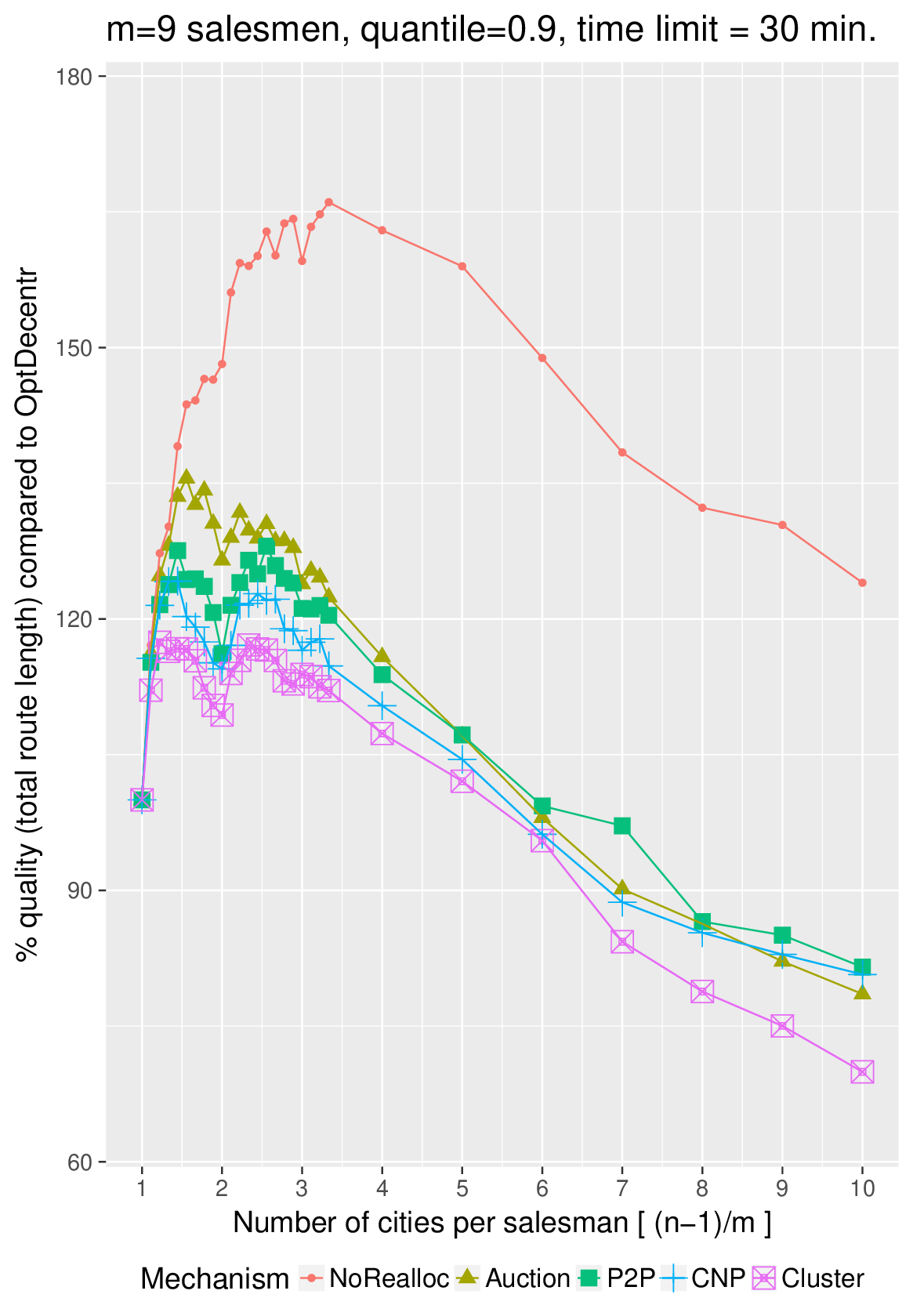}
	\includegraphics[width=6.268cm,height=10.5cm]{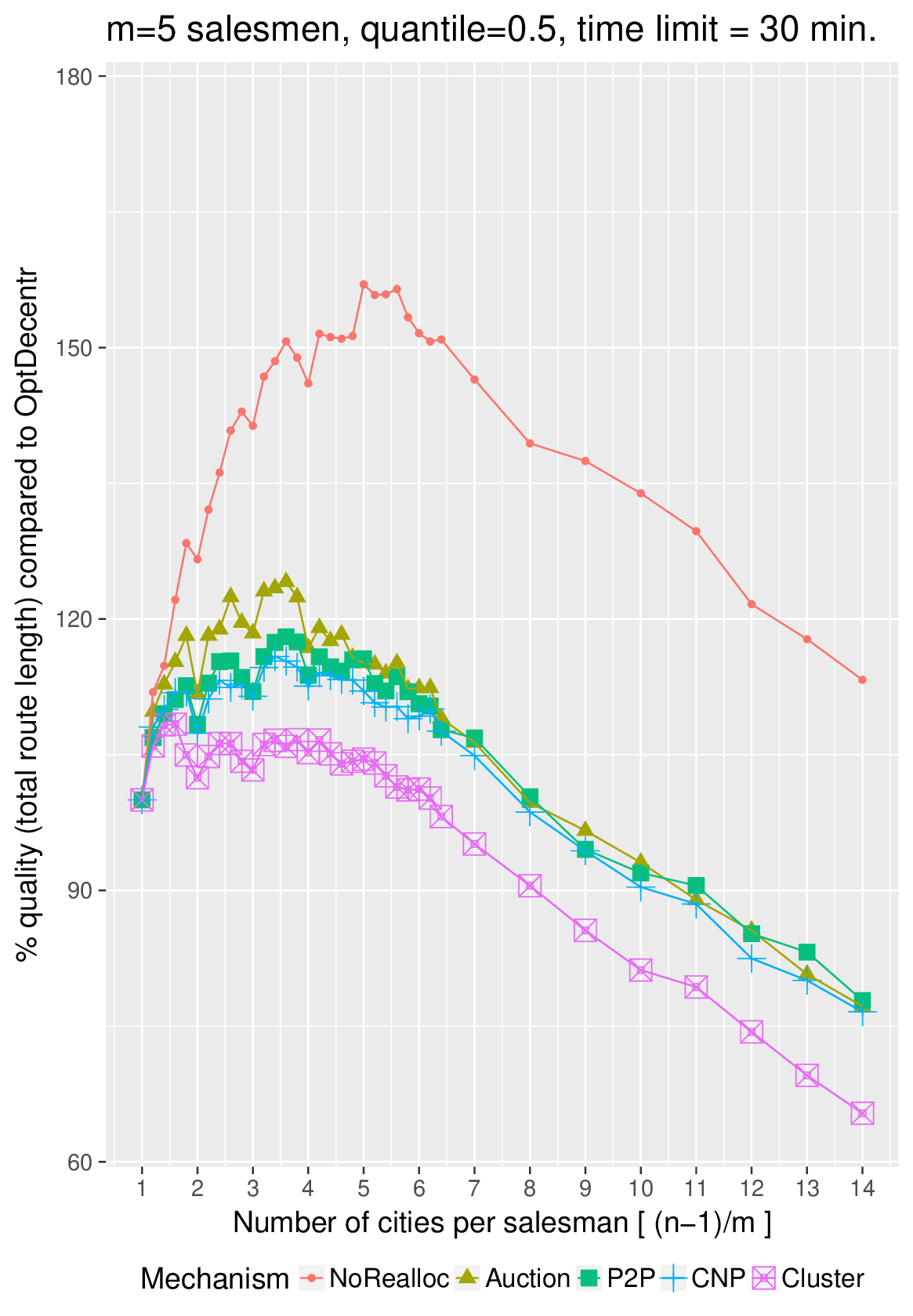}
	\includegraphics[width=6.268cm,height=10.5cm]{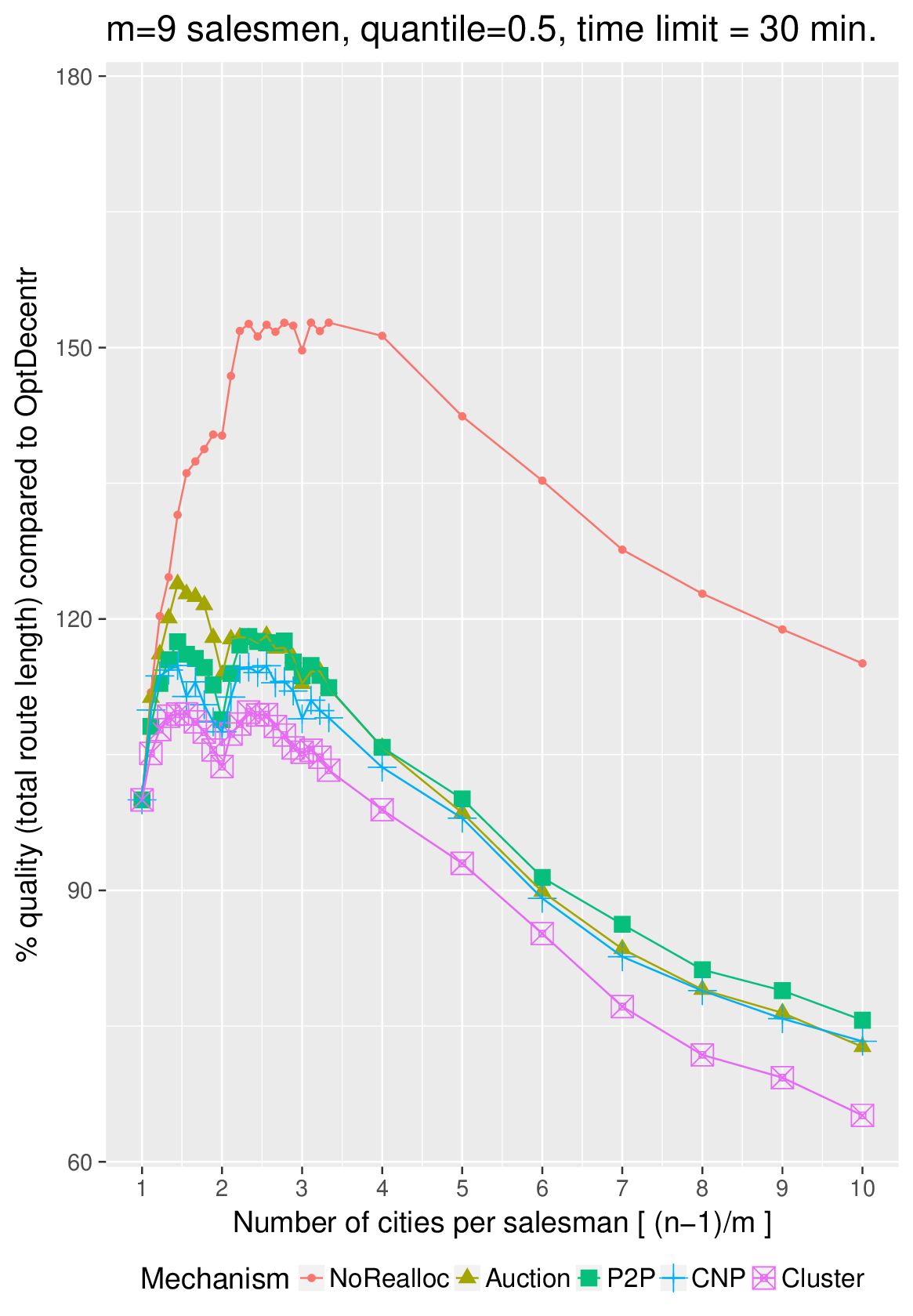}
	\caption{Ratios of quality compared to OptDecentr for $m=5$ (left) and $m=9$ (right) salesmen when the time limit is 30 minutes.}
	\label{fig-ratios-30minutes-OptDecentr}
	\end{center}
\end{figure}

\begin{figure}
	\begin{center}
	\includegraphics[width=6.268cm,height=10.5cm]{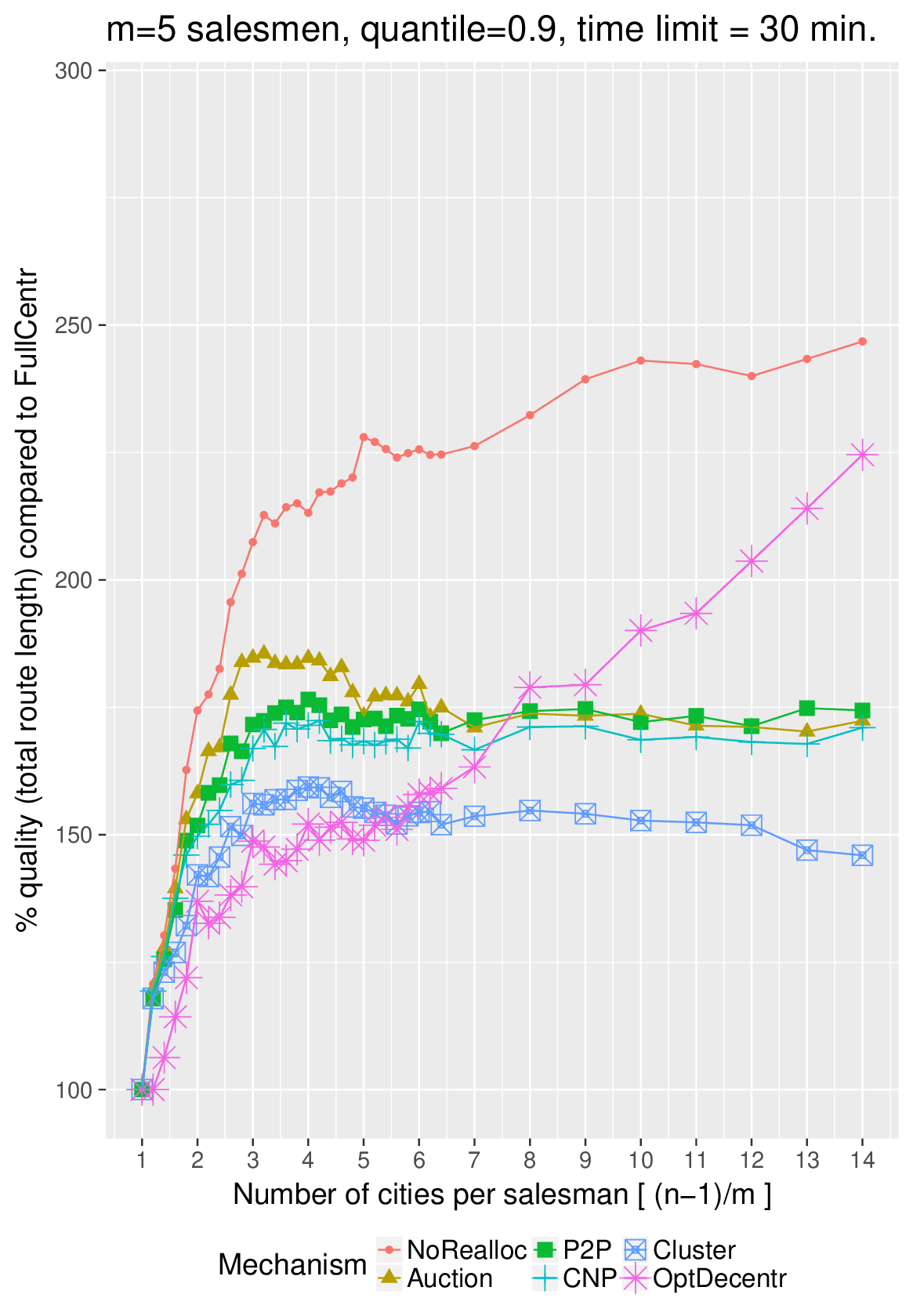}
	\includegraphics[width=6.268cm,height=10.5cm]{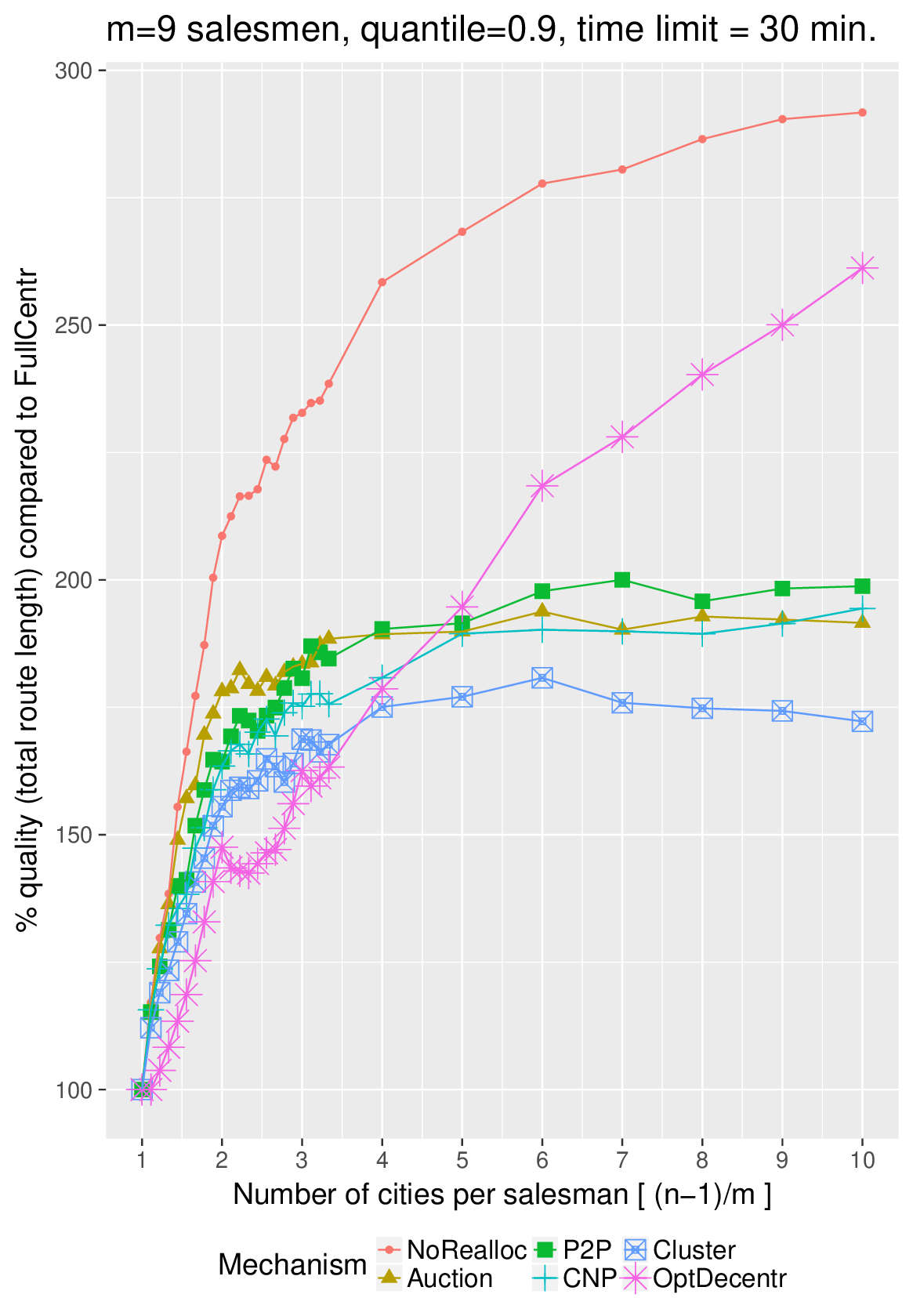}
	\includegraphics[width=6.268cm,height=10.5cm]{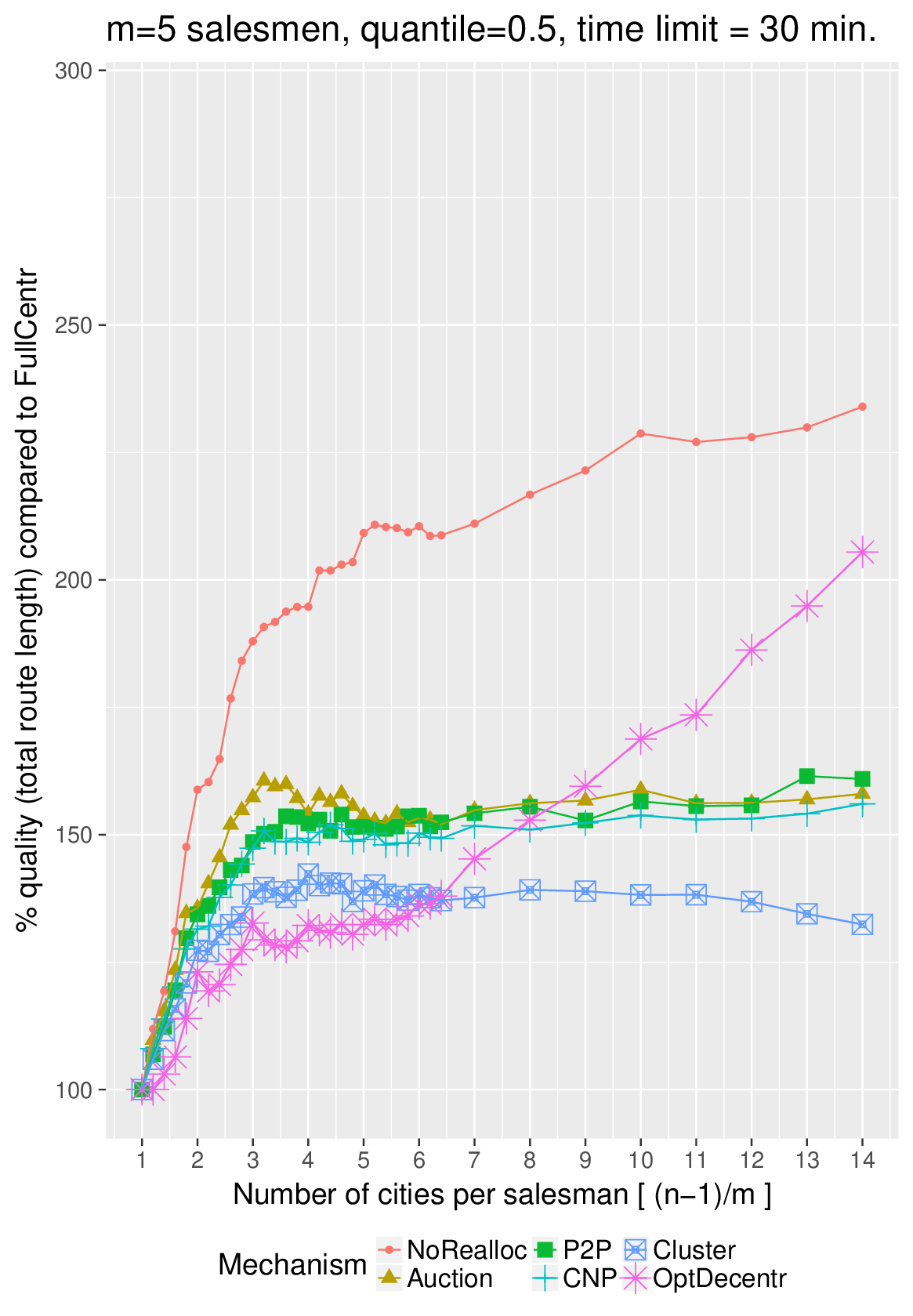}
	\includegraphics[width=6.268cm,height=10.5cm]{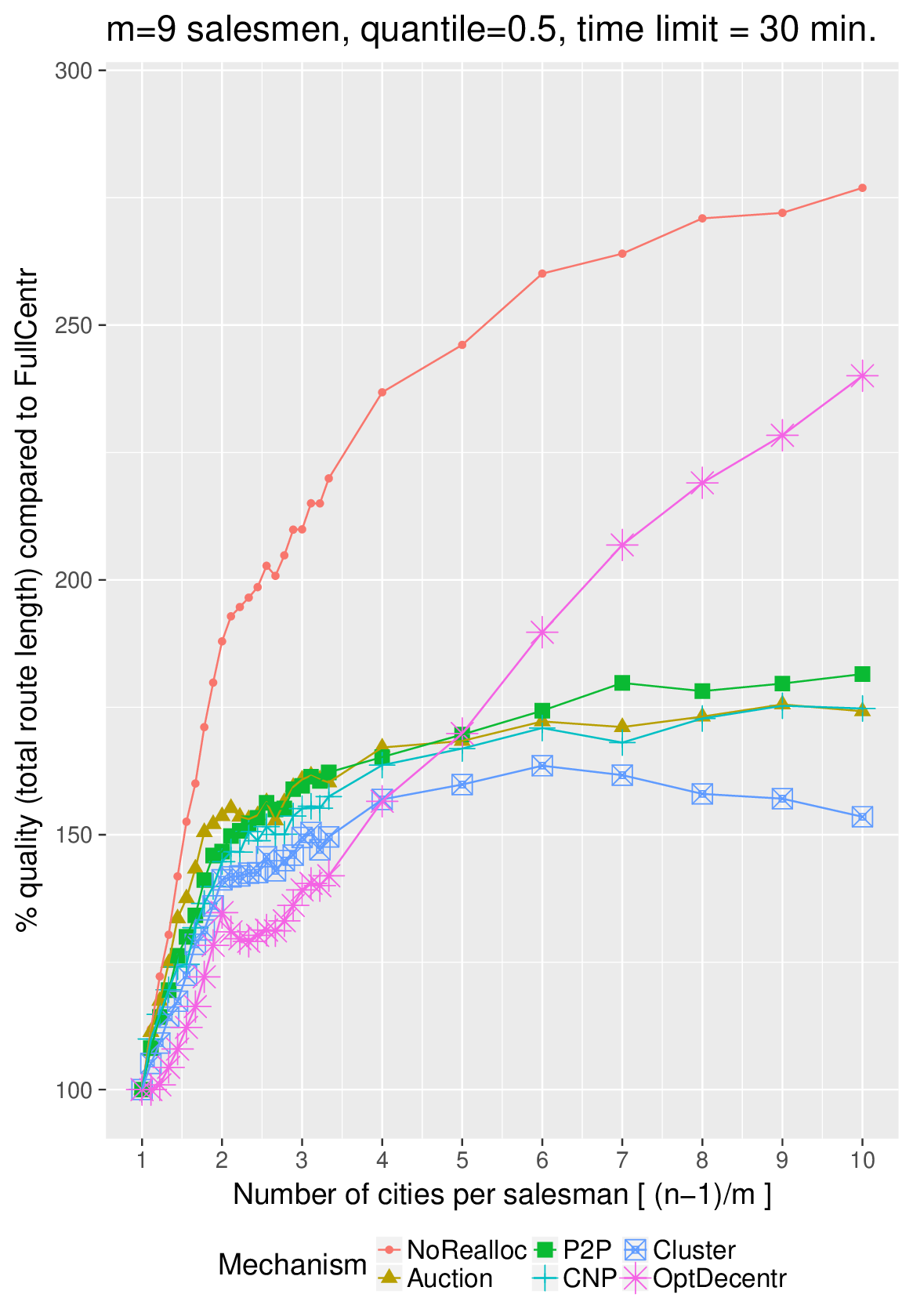}
	\caption{Ratios of quality compared to FullCentr for $m=5$ (left) and $m=9$ (right) salesmen when the time limit is 30 minutes.}
	\label{fig-ratios-30minutes-FullCentr}
	\end{center}
\end{figure}

In order to obtain figures with much larger numbers of cities $n$, we limit the computation time to 30 minutes. (Hence, we only show the ratios of total route lengths since the equivalent of Figure \ref{fig-boxplots-infinity-TOTcmpT} would only show that this time limit is respected.)
In both Figures \ref{fig-ratios-infinity-OptDecentr} and \ref{fig-ratios-infinity-FullCentr}, the real-life duration of the operation of the mechanisms inferred from sequential simulations, as explained in Section \ref{duration}, was computer after the end of the simulations (offline).
In order to obtain more data, we have modified our AnyLogic models such that this computation is done throughout the simulation (online) by updating {\tt get\_Main()\-.re\-main\-ing\-ComputationTime} which is shared by all salesmen, and added {\tt cplex.setParam(IloCplex.Param.TimeLi\-mit, get\_Main()\-.re\-main\-ing\-ComputationTime/1000)} in order to make CPLEX stop before or at the time limit. In Figures \ref{fig-ratios-30minutes-OptDecentr} and \ref{fig-ratios-30minutes-FullCentr}, this time limit is set to 30 minutes in parallel simulations.
That is, \ac{DO} mechanisms may run much longer sequential simulations in AnyLogic, but they cannot last more than 30 minutes when we infer what they would last in reality.

As a consequence, \sf{OptDecentr} and \sf{FullCentr} find the optimal solution of their respective version of \ac{MTSP} in Figures \ref{fig-ratios-infinity-OptDecentr} and \ref{fig-ratios-infinity-FullCentr} because we stopped the experimentation as soon as the time limit is reached. Hence, it was not possible to have a point below 100\%. On the contrary, Figure \ref{fig-ratios-30minutes-OptDecentr} shows points below 100\% for high values of $n$ when the simulation stops before CPLEX has found the optimal solution of \sf{OptDecentr}. (Figure \ref{fig-ratios-30minutes-FullCentr} has no points below 100\% because \sf{FullCentr} is much quicker to optimiser.)

The most salient points to observe in Figures \ref{fig-ratios-30minutes-OptDecentr} and \ref{fig-ratios-30minutes-FullCentr} are as follows:
\begin{itemize}
	\item \emph{The cost of selfishness is about 30\% higher for $m=5$ and 60\% higher for $m=9$ than the cost in the traditional \ac{MTSP}}: In order to see this, we compare \sf{FullCentr} (\ac{MTSP} without our constraint modelling selfishness) and \sf{OptDecentr} (benchmark showing the best results that \sf{Cluster}, \sf{CNP}, \sf{P2P} and \sf{Auction} could find):
	\begin{itemize}

		\item \emph{For the instances of small size} ($i.e.$, less than 6 cities per salesman for $m=5$ in the lower left graph in Figure \ref{fig-ratios-30minutes-FullCentr}, and less than 4 cities per salesman for $m=9$ in lower right), \sf{OptDecentr} finds the best \ac{DO} solution because the time limit of 30 minutes does not stop this mechanism too early.

		\item \emph{For larger instances}, \sf{OptDecentr} does not have enough time to find the optimum. Therefore, we look at the quality of \sf{Cluster} instead, because it is the best \ac{DO} mechanism. In other words, we use \sf{Cluster} in order to infer the quality that \sf{OptDecentr} would obtain without the time limit for such larger instances. In fact, \sf{OptDecentr} would find the optimal solution of our modified \ac{MTSP} if there were no time limit, hence we assume that the quality of the solution found by \sf{Cluster} is an upper bound of a perfect \ac{DO} mechanism.
	\end{itemize}
Since the median route length of \sf{Auction} reaches a plateau at about 130\% (respectively, 160\%) the length of \sf{FullCentr} for $m=5$ (respectively, $m=9$), we conclude that these figures are upper bounds of the median route length of \sf{OptDecentr}. As a conclusion, the modelling of selfishness in our modified \ac{MTSP} increases the total route length by 30\% (respectively, 60\%) for $m=5$ (respectively, $m=9$) in comparison with the traditional \ac{MTSP}.

	\item \emph{(De-)centralisation seems to have a negligible impact on the quality of the solution:} \sf{CNP}, \sf{Auction} and \sf{P2P} have very close results in Figure \ref{fig-ratios-30minutes-FullCentr} when $n$ is large, while the second of these mechanisms has a \ac{CA}, $i.e.,$ an auctioneer. The next bullet point indicates that this may due to the fact that this \ac{CA} is not coercive.

	\item \emph{Coercion seems to improve the quality of the solution:} Both Figures \ref{fig-ratios-30minutes-OptDecentr} and \ref{fig-ratios-30minutes-FullCentr} suggest that the more coercive the \ac{CA} in a mechanism, the better the quality of the solution found by this mechanism:
	\begin{itemize}
		\item \emph{Ranking of mechanisms by quality of the solution}: If we write $>$ for ``is more efficient than'', then we can see in Figures \ref{fig-ratios-30minutes-OptDecentr} and \ref{fig-ratios-30minutes-FullCentr} that \sf{FullCentr} $>$ \sf{OptDecentr} $>$ \sf{Cluster} $>$ \sf{Auction}.
		\item \emph{Ranking of mechanisms by coercion level of their \ac{CA}}: The same order applies to the coercion level of the \ac{CA} in these four mechanisms:
		\begin{enumerate}
			\item \sf{FullCentr} has a more coercive \ac{CA} than \sf{OptDecentr} because she ignores the selfishness of the salesmen.
			\item The \ac{CA} in \sf{OptDecentr} is more coercive in \sf{Cluster} because she controls both allocation and routing.
			\item The \ac{CA} in \sf{Cluster} requires to know of all cities while, as said in the previous bullet point,  she is just a mediator in \sf{Auction}.
		\end{enumerate}
	\end{itemize}
\end{itemize}

Shortly, Figures \ref{fig-ratios-30minutes-OptDecentr} and \ref{fig-ratios-30minutes-FullCentr} for high value of $n$ suggest the following ranking (from most to least efficient): \sf{FullCentr} $>$ \sf{OptDecentr} $>$ \sf{Cluster} $>$ (\sf{Auction} $\approx$ \sf{CNP} $\approx$ \sf{P2P}) $>$ \sf{NoRealloc}.

\section{Discussion}
\label{section-discussion}

This discussion puts into perspective this article with regard to the general question of the price of the selfishness in \ac{DO}. This presentation is carried out according to our contributions:

\subsection{Conceptual contributions}

This article focuses on one way to include \ac{DO} features in human organisations -- namely, the selfishness of the salesmen -- into the traditional \ac{MTSP} addressed by \ac{CO}. However, please notice that other pairs of \ac{CO}/\ac{DO} problems are possible. These other possibilities may be summarised as follows: 
\begin{itemize}
	\item \emph{Relax Assumption \sf{Hyp.~3}}: As aforementioned, \sf{Hyp.~3} may be relaxed by replacing the constraint of 1-1 exchanges by $n$-$n$ exchanges. Clearly, the cases with $n>1$ may find better exchanges but a combinatorial number of possible exchanges would have to be considered by salesmen.
	\item \emph{Modify the objective function in \ac{MTSP}}: As shown in Subsection \ref{modifiedMTSP}, we choose to keep the same social welfare as the objective function in the traditional \ac{MTSP}, which eventually result in adding the constraint of 1-1 exchanges with an initial endowment of cities.

	Instead of such a modification of the constraints, we could have modified the objective function. For example, we could have i) given a value $v_{ik}$ to each City $i$ for each Salesman $k$ (for instance, Salesman 1 earns $v_{11}$=\euro 3 for visiting City 1 and $v_{21}$=\euro 4 for City 2 while Salesman 2 receives $v_{12}$=\euro 6 and $v_{22}$=\euro 5 respectively) and ii) modified the objective function to make a trade-off between the distance travelled and the value earned for visiting cities (for example, this distance is transformed into a cost of fuel and this cost is subtracted from the money earned in the cities).
With such a utility function, the salesmen would agree to increase the number of their cities when their value is high enough.

	Other examples of modifications rely on the fact that our modified \ac{MTSP} relies on the utilitarian social welfare ($i.e.,$ sum of individual utilities), but other mappings of individual utilities to a social welfare have been proposed ($e.g.,$ maximum, minimum or product of individual utilities).

	\item \emph{Derive a \ac{CO} problem from a \ac{DO} one}: Instead of adding the selfishness of the decision makers to a \ac{CO} problem, other work in the literature do the contrary. For instance, \cite{sallez10} compare the dynamic allocation and routing in a real flexible manufacturing system managed by a \ac{DO} mechanism to a \ac{CO} benchmark which does not take all the constraints into account because of the combinatorial explosion.
	\item \emph{Change who make decisions}: We assume that salesmen fight for cities, but the opposite is also possible. It is even possible that both salesmen and cities make decisions.
\end{itemize}

Finally, we observe that \ac{CO} can estimate the performance of \ac{DO}. In fact, \sf{OptDecentr} implements \ac{CO} in order to find the best solution of our modified \ac{MTSP}, and this solution is a benchmark for \sf{Cluster}, \sf{CNP}, \sf{P2P} and \sf{Auction}.

\subsection{Technical contributions}
Our main technical contribution is the implementation of various mechanisms solving our modified \ac{MTSP}.
It is interesting to notice that we have implemented one mechanism per organisation, but other mechanisms are possible for all the organisations in Figure \ref{organisations.eps}.
For example, \sf{Cluster} uses the \ac{MILP} formulation proposed by \cite{rao71}, but others exist.\footnote{We have also tested Mechanism \sf{Cluster} with the formulation by \cite{saglam06} to which we added our constraint for 1-1 exchanges. Our experimentation (not presented in this article) shows similar performance than the formulation by \cite{rao71} described in Paragraph \ref{Cluster}, even though the obtained clusters sometimes differ.}

We have chosen to measure the computation time of CPLEX only.
On the one hand, we first started to program our mechanisms with a \ac{MILP} solver written in Java, namely ojAlgo v40 (\url{http://www.ojalgo.org} in order to have only pure Java in AnyLogic. We stopped this and turned to CPLEX because it is a recognised benchmark while we know little about the performance of ojAlgo.
On the other hand, we also first thought about using various tools and environments (Concorde, k-means, etc.), next use benchmarks to compare their computation times. Unfortunately, no recognised benchmarks exist and contradicting information may even be found, such as ``C runs faster than Java because it operates on a lower level'' and ``Java runs faster than C because its virtual machine adapts the program to the computer''.

\subsection{Numerical contributions}

Besides the data shown in Section \ref{experimentation}, especially the bullet points in Subsection \ref{result-30minutes}, our results are robust because each point in Figures \ref{fig-ratios-infinity-OptDecentr}, \ref{fig-ratios-infinity-FullCentr}, \ref{fig-ratios-30minutes-OptDecentr} and \ref{fig-ratios-30minutes-FullCentr} represent a decile calculated on 130 instances. Of course, our conclusions at the end of Subsection \ref{result-30minutes} may turn out to be wrong with other instances, as well as with other hypotheses about how to modify \ac{MTSP} in order to take the selfishness of the salesmen into account.

\section{Conclusion}
\label{Conclusion}

This article compares more or less centralised organisations in order to quantify the cost of having selfish decision makers. For that purpose, our first contribution is to introduce a same problem with features for both \acf{CO} and \acf{DO}. We address this issue by constraining the \acf{MTSP} with both 1-1 exchanges of cities an an initial endowment.
Our second contribution is to be the first article comparing five decision organisations to solve a same joint allocation and routing problem, while the few other similar comparisons only address two organisations.
Our third contribution is the quantification of the cost of decentralising the making of decisions. We think that the most interesting result is the fact that \ac{DO} ($i.e.,$ our modified \ac{MTSP}) has a median total route length which reaches a plateau $\approx$30\% (respectively, $\approx$60\%) longer than \ac{CO} ($i.e.,$ the traditional \ac{MTSP}) when there are five (respectively, nine) salesmen and many cities. This stabilisation was hoped but unpredictable without experimentation. We also notice that the coercion level of \ac{CA} seems to impact much more on the quality of the solution than the level of centralisation of a mechanism.

As future work, we plan to study how our model of selfishness impacts on the efficiency of the mechanisms. For that purpose, we will use the model of preferences detailed in the discussion section in which the salesmen make a trade-off between the value of cities and the distance travelled, and then adapt our mechanisms to this other variant of \ac{MTSP}. 

\section*{Acknowledgement}

We would like to thank the department of Industrial Engineering of INSA-Lyon, France
for the use of the computers in one of its student laboratories in order to obtain all the results in this article.

\bibliographystyle{apalike} 
\bibliography{biblio,biblio2,biblio3}

\end{document}